\documentclass[draftcls,11pt,onecolumn]{IEEEtran}
\usepackage{amsmath,amssymb,graphicx}

\title{A Novel Approach for Compression of Images Captured using Bayer Color Filter Arrays\thanks{This work was supported in part by NASA under grant AIST-0122-005.}}
\author{Sang-Yong~Lee, ~\IEEEmembership{Member,~IEEE,} and~Antonio~Ortega,~\IEEEmembership{Fellow,~IEEE}\\
Ming Hsieh Department of Electrical Engineering\\
         Signal and Image Processing Institute\\
         University of Southern California\\
         antonio.ortega@sipi.usc.edu}

\begin{document}

\maketitle \IEEEpeerreviewmaketitle

\begin{abstract}
We propose a new approach for image compression in digital
cameras, where the goal is to achieve better quality at a given
rate by using the characteristics of a Bayer color filter array.
Most digital cameras produce color images by using a single CCD
plate, so that each pixel in an image has only one color component
and therefore an interpolation method is needed to produce a full
color image. After the image processing stage, in order to reduce
the memory requirements of the camera, a lossless or lossy
compression stage often follows. But in this scheme, before
decreasing redundancy through compression, redundancy is increased
in an interpolation stage. In order to avoid increasing the
redundancy before compression, we propose algorithms for image
compression in which the order of the compression and
interpolation stages is reversed. We introduce image transform
algorithms, since uninterpolated images cannot be directly
compressed with general image coders. The simulation results show
that our algorithm outperforms conventional methods with various
color interpolation methods in a wide range of compression ratios.
Our proposed algorithm provides not only better quality but also
lower encoding complexity because the amount of luminance data
used is only half of that in conventional methods.
\end{abstract}


\section{Introduction}
\label{sec:intro} \vspace{0.0cm} \PARstart{D}{igital} cameras use
image-processing tools, e.g., interpolation techniques, such as
those used in analog camcorders, in order to achieve good quality
images. One big difference between digital cameras and analog
camcorders is that digital cameras store digital data in flash
memories. Thanks to storing digital data, functionalities such as
image editing and enhancement can be added. But the price of flash
memories is still very high, so that low image bit-rates are
required in order to enable storage of large number of images at a
reasonable cost.
To achieve this, most digital cameras use lossy compression
schemes like JPEG~\cite{JPEGBook} to store the images.

\begin{figure}[tb]
\centering \includegraphics[width=3.5cm]{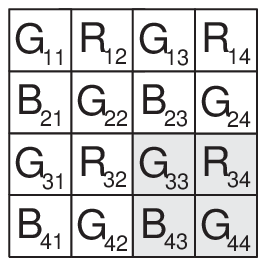} \caption{Bayer
color filter array. Each letter indicates the position of a
different color filter. R, G and B are for Red, Green and Blue,
respectively. The gray block indicates 2 by 2 repeating pattern.}
\label{fig:1}
\end{figure}

In this paper, we focus on the color interpolation process that is
used in many cameras, and specifically on how this interpolation
should be taken into consideration when designing image
compression for digital cameras. In order to produce full color
images, most digital cameras place color filters on monochrome
sensors. While some high-end digital cameras use three CCD plates
to get full color images, where each plate takes one color
component, most digital cameras use a single CCD plate, with
several different color filters, and produce full color images by
using an interpolation technique. Although there are several
different color filter arrays (CFA)~\cite{Yamada00SSC}
\cite{Yamanaka77P}, in this paper, we focus on the Bayer CFA which
is most widely used in digital cameras. The Bayer CFA, as shown in
Fig.~\ref{fig:1}, uses 2 by 2 repeating patterns (RP) in which
there are two green pixels, one red and one blue. There is only
one color component in each pixel, so the other two color
components for a given pixel have to be interpolated using
neighboring pixel information. For example, in a bilinear
interpolation method, the red (blue) color component on a green
pixel in Fig.~\ref{fig:1} is produced by the average value of two
adjacent red (blue) pixels. Although there are several possible
interpolation algorithms~\cite{Ramanath02JEI}
\cite{Longere02PIEEE} \cite{Trussell02IP} \cite{Adams98ICIP}
\cite{Kimmel99IP} \cite{Kehtarnavaz03JEI}, it is clear that from
an information theoretic viewpoint they all result in an increase
of
redundancy. 

\begin{figure}[tb]
\centering
\begin{minipage}[b]{1.0\linewidth}
\centering
\includegraphics[width=8.8cm]{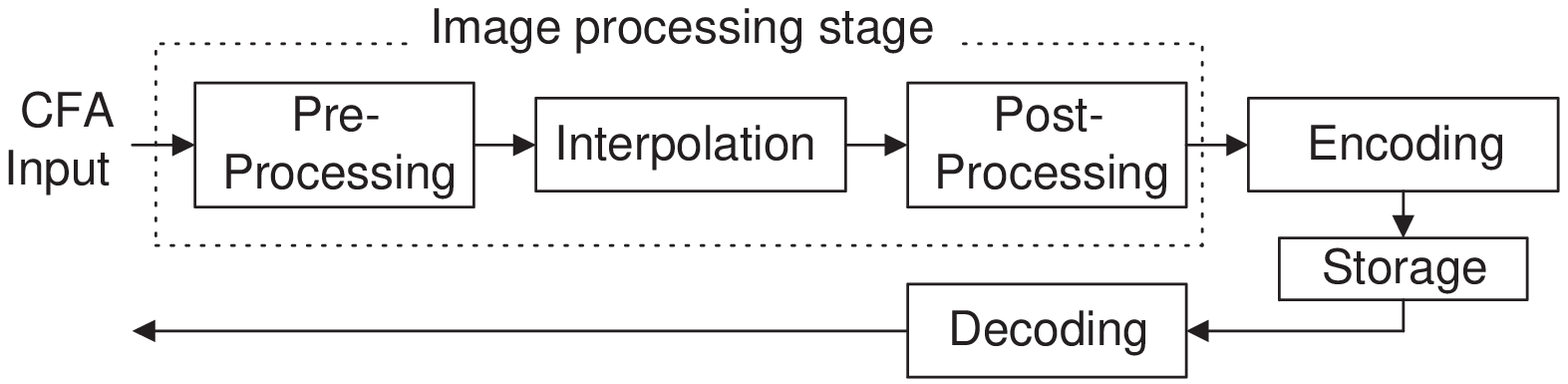}\\
  \vspace{0.0cm}
 \centerline{(a)}\smallskip
\end{minipage}
\begin{minipage}[b]{1.0\linewidth}
\centering
\includegraphics[width=8.8cm]{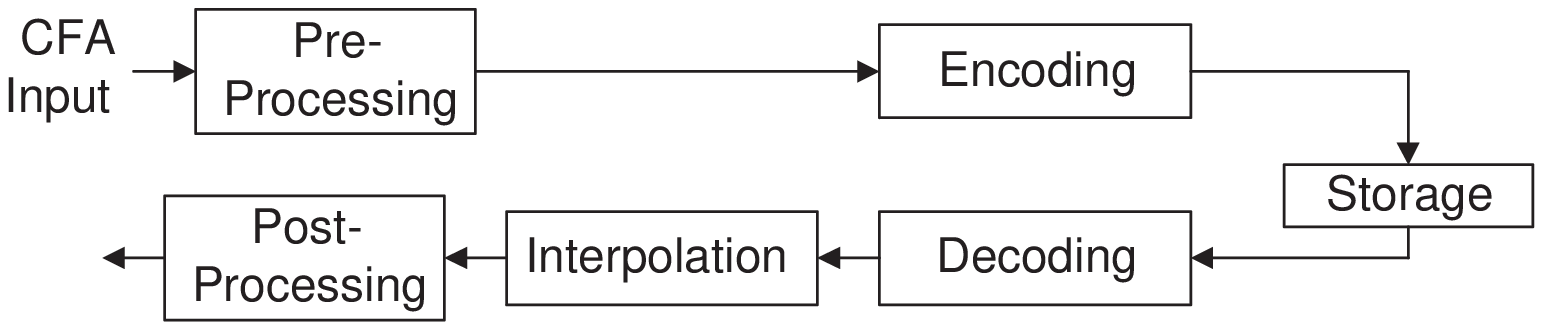}\\
  \vspace{0.0cm}
\centerline{(b)}
\end{minipage}
\caption{Block diagrams of (a) the conventional method and (b) the
proposed method. In (a) an image processing stage is followed by a
compression stage. In (b) interpolation and post-processing in an
image processing stage are done after compression and
decompression.} \label{fig:diagram} \vspace{0.0cm}
\end{figure} 
In a conventional method, as shown in Fig.~\ref{fig:diagram} (a),
after finishing the image processing stage, a lossless or lossy
image compression algorithm is used before storing the image.
Although in theory one could achieve the same compression with or
without the interpolation, to do so would require exploiting the
specific characteristics of the interpolation technique within the
compression algorithm, which is clearly not easy to achieve if we
wish to use a standard compliant compression method without
modification. For this reason, in this paper we propose image
transform algorithms to encode the image {\em before}
interpolation, so that interpolation is performed only after
decoding. We call this approach Interpolation after Decoding
(IAD), see Fig.~\ref{fig:diagram} (b). There are some other
functions, such as white balancing and color correction, that are
performed in the image processing stage. These are shown as pre-
and post-processings in the figure. In our proposed approach less
data needs to be encoded, since only one color is available at
each pixel position before interpolation. The main challenge is
then how to organize the available data for encoding in order to
best exploit the spatial redundancy for compression.

Methods to use increase image quality using the redundancy of
interpolation in post- and pre-processing stages and using encoder
characteristics during interpolation have been studied.
In~\cite{Herley98ICIP}, under the assumption of a fixed
interpolation algorithm, the quantization noise is reduced by
using an iterative method that incorporates information about the
interpolation algorithm. By contrast our approach assumes only a
specific CFA and can operate with any interpolation technique.
The main difference is that our algorithms compress
non-interpolated images without introducing the redundancy of
interpolation, whereas the algorithm in~\cite{Herley98ICIP}
improves image quality by exploiting the redundancy of
interpolation.

In~\cite{Baharav02SPIE}, under a given compression method,
minimizing error between a decoded full color source and a decoded
image after interpolation is studied, but the complexity required
is too high to use in digital cameras. In~\cite{Koh03ICIP}, as a
modified approach of our previous work in~\cite{Lee01ICIP}, CFA
data compression with different format conversion is proposed but
the method is limited to work with bilinear interpolation. Since
in our algorithm, interpolation is not involved in encoding and
decoding processes, the algorithm itself is independent from
interpolation methods. Also, interpolation is done on the decoded
pixels and so any interpolation method which is not sensitive to
the coding error can be applied.

As an image coder, JPEG is widely used in digital cameras because
it is relatively simple and provides good performance, especially
when the compression ratio is low. JPEG is a block discrete cosine
transform (DCT) based coder and the blocking artifacts can become
severe as the compression ratio becomes higher. Discrete wavelet
transform (DWT) based coders such as EZW~\cite{Shapiro93SP},
SPIHT~\cite{Said96CSVT} and EBCOT~\cite{Taubman00IP} (adopted in
JPEG2000~\cite{JPEG2000Book}) are also used as image coders. A DWT
based coder does not produce blocking artifacts and it provides
good performance at high compression ratios. In this paper, we use
JPEG and SPIHT as representative of the DCT and DWT based
approaches, respectively. Our proposed algorithms are tested under
both of these coding techniques. Although we focus on standard
compression methods, interpolation aware compression methods can
provide better performance especially when the interpolation
method used is sensitive to the coding error of standard
compression methods.

In this paper, extending our previous work in~\cite{Lee01ICIP}, we
propose several different algorithms to transform the
non-interpolated images before compression. We provide performance
results of the proposed algorithms with different coders (JPEG and
SPIHT) and interpolation methods (bilinear and adaptive
interpolation). Also, using a simple example based on
one-dimensional data, we propose an analysis to provide some
intuition about why our approach outperforms conventional
compression-after-interpolation (CAI) methods. In our problem an
original full color image is not available, since cameras are
assumed to capture images with single color pixels. Thus, for the
purpose of comparison we use as a reference a full color image
obtained by interpolating the original (uncompressed) captured
image. Therefore our problem will be to find coding schemes that
are optimized in terms of minimizing the error with respect to
that original interpolated image. The experimental results show
that the proposed algorithms outperform the conventional method in
the full range of compression ratios for JPEG coding with bilinear
interpolation and up to 20 : 1 or 40 : 1 compression ratio for
SPIHT coding depending on the interpolation methods used. Thus, in
both cases, our proposed techniques are superior in the range of
compression ratios that are used in practical digital cameras
(i.e., those corresponding to high quality images).

This paper is organized as follows: in Section~\ref{sec:toy}, the
theoretical rate distortion performance of the CAI and IAD
approaches is analyzed by using a 1-D sequence and DPCM encoding.
Proposed image transform algorithms are addressed in
Section~\ref{sec:algorithm}. Experimental results are provided as
demonstration of the validity of our algorithm in
Sections~\ref{sec:capt_results} and~\ref{sec:adaptive}. Finally,
the conclusion of this work is in Section
\ref{sec:capt_conclusion}.

\section{Performance comparison using one dimensional sources}
\label{sec:toy} The main difference between the CAI and IAD
methods is the order of compression and interpolation. In this
section we propose an analysis to provide some intuition about why
an IAD method can theoretically outperform a CAI method by
considering differential pulse code modulation (DPCM) compression
of a one dimensional first order autoregressive (AR) process. DPCM
exploits the correlation between two adjacent pixels to reduce the
residual energy coded, whereas transform coding used in standard
image coding exploits spacial correlation to pack a large fraction
of its total energy in relatively few transform coefficients.
Therefore DPCM compression has similar coding process (including
decorrelation, quantization and entropy coding) to standard image
coding, although the performance of decorrelation in DPCM is
weaker than that in transform coding.

First, we compare the R-D performance of DPCM and DPCM after
interpolation (DPCMI). Then, we show that the IAD method
outperforms the CAI method under the following assumptions: (i)
DPCM coding is used, (ii) the interpolated sequence is divided
into two sub-sequences during coding and (iii) the distortion is
measured after interpolation.

Although open loop DPCM is not generally used due to the error
propagation in the decoded sequence, its analysis is easier, given
that the difference sequence has an explicit theoretical R-D curve
when the source is a Gaussian AR process. 
\begin{figure}[tb]
\centering
 \begin{tabular}{cc}
\includegraphics[width=4.8cm]{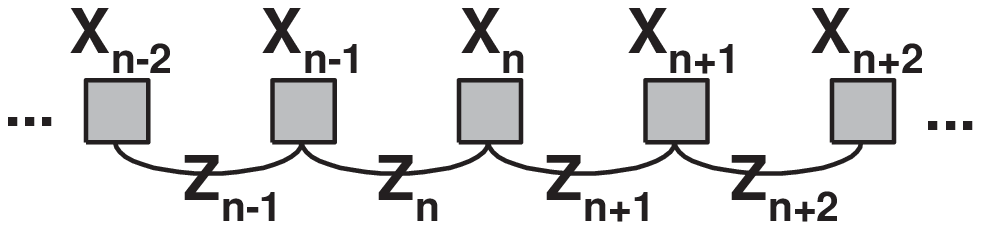} &
\includegraphics[width=4.8cm]{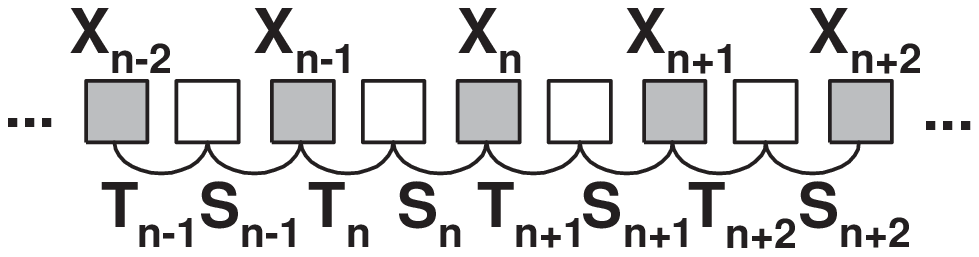}
\\ \vspace{0.2cm}
   (a) & (b)\\
  \end{tabular}
 \caption{Gray and white boxes indicate original and interpolated
samples, respectively. In (a), $\{Z_n\}$ is a differential
sequence of the original sequence taken from sensors and in (b),
$\{T_n\}$ and $\{S_n\}$ indicate a differential sequence of the
interpolated sequence.} \label{fig:DPCM}
\end{figure} 
Therefore we consider open loop DPCM of one dimensional first
order zero mean Gaussian AR processes. Let
\setlength{\arraycolsep}{0.0cm}
\begin{equation}\label{eq:AR}
 X_n{}={}\rho X_{n-1}{}+{}W_n, ~ n = 1, 2, \cdots ,
\end{equation}
\setlength{\arraycolsep}{5pt}
denote the process, where $\{W_n\}$
is a zero-mean sequence of independent and identically distributed
random variables and $W_n \sim N(0,\sigma_W^2)$, and $\rho$ is the
correlation coefficient ($0\leq \rho < 1$). Then from the
probability distribution of $W_n$, the probability distribution of
$X_n$ is $N(0, \sigma_W^2 /(1-\rho^2))$. We assume that the
initial state $X_0$ is given and we are interested in the source
outputs for $~n \geq 1$.

We define the differential sequence of $\{X_n\}$ as $\{Z_n\}$
\setlength{\arraycolsep}{0.0em}
\begin{equation}\label{eq:DPCM}
 Z_n \triangleq X_n - X_{n-1} \\
 {}={} (\rho -1)X_{n-1}{}+{}W_n~.
\end{equation}
\setlength{\arraycolsep}{5pt}
Since $X_{n-1}$ and $W_n$ are
independent, $Z_n$ also has Gaussian distribution ($Z_n \sim
N(0,~2 \sigma_W^2 /(1+\rho))$).

The rate distortion (R-D) function for a Gaussian source with mean
square error (MSE) distortion can be written in closed
form~\cite{cover91info} and therefore the R-D function of ${Z_n}$
is
\begin{equation}\label{eq:RD_1}
 R_1(D) = \frac{1}{2} \log_2 (\frac{2 \sigma_W^2}{(1+\rho)D}),~~~~ \text{for }0 \leq D <\frac{2
 \sigma_W^2}{1+\rho}~,
\end{equation}
where $D$ denotes average distortion of the data coded. The
distortion of interpolated pixels is addressed in the last part of
this section.

Next, we double the number of samples by using a linear
interpolation method and define this new sequence $Y_n$ as :
\begin{equation}\label{eq:INT}
\begin{cases}
 Y_{2n} \triangleq X_n, \\
 Y_{2n+1}   \triangleq (X_n + X_{n+1})/2.
 \end{cases}
\end{equation}
From this sequence, as shown in Fig.~\ref{fig:DPCM} (b), two
differential sequences $\{T_n\}$ and $\{S_n\}$ can be defined as
\begin{equation}\label{eq:INT2}
\begin{cases}
 T_n \triangleq Y_{2n-1}-Y_{2n-2} \\
 S_n \triangleq Y_{2n}- Y_{2n-1}.
 \end{cases}
\end{equation}
Note that because of the chosen interpolation mechanism, $T_n$ is
identical to $S_n$ (i.e., $T_n = S_n = (X_n - X_{n-1})/2$) and the
probability distribution of $T_n$ (or $S_n$) is
$N(0,\sigma_W^2/2(1+\rho))$. Since both $\{T_n\}$ and $\{S_n\}$
are Gaussian sources, their R-D functions are :
\begin{equation}\label{eq:RD_T}
 R_T(D) = R_S(D) = \frac{1}{2} \log_2
 (\frac{\sigma_W^2}{2(1+\rho)D}),
 ~\text{for } 0 \leq D < \frac{\sigma_W^2}{2(1+\rho)}~.
\end{equation}
In this example, since $S_n$ and $T_n$ are same, there is no need
to encode $S_n$ if $T_n$ is available. However, in our original
problem, the difference of neighboring pixels of interpolated
images is not same since more than two pixels are involved in 2-D
interpolation. Also, a standard compliant compression algorithm
cannot employ additional information related to interpolation.
Therefore, although $S_i$ and $T_j$ can be spatially correlated,
we assume that $S_i$ and $T_j$ will be encoded independently for
all $i$ and $j$. Then the rate distortion function of DPCM for
$Y_n$, i.e., the R-D function for the DPCMI approach, will be :
\begin{equation}\label{eq:RD_2}
 R_2(D) = R_T(D) + R_S(D) = \log_2 (\frac{\sigma_W^2}{2(1+\rho)D}),
 ~\text{for } 0 \leq D < \frac{\sigma_W^2}{2(1+\rho)}~.
\end{equation}
\begin{figure}[tb]
\centering
\includegraphics[width=11cm]{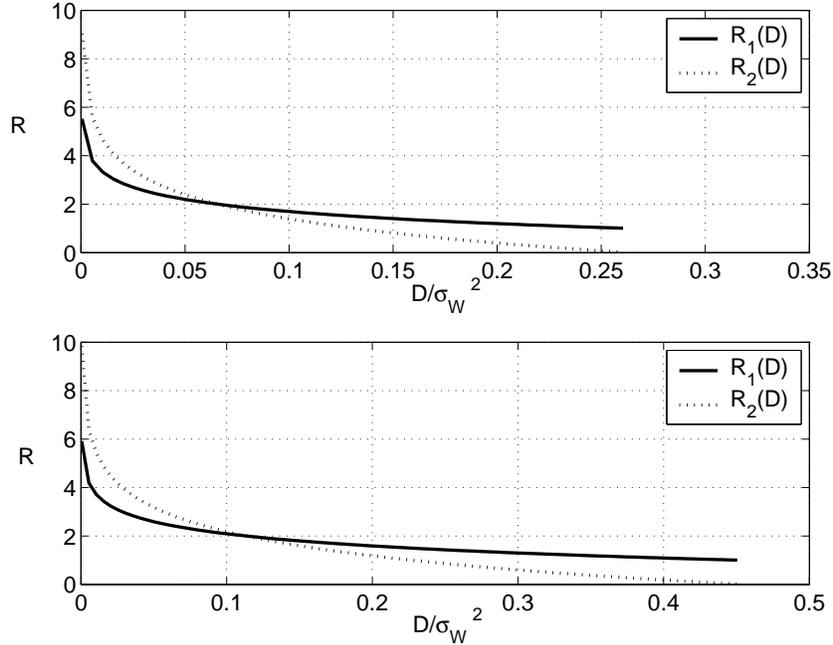}
\caption{The upper (lower) graph is for $\rho = 0.9$
($\rho = 0.1$). Solid lines indicate the R-D curve of the
differential sequence ($\{Z_n\}$) and dotted lines indicate the
R-D curve of the differential sequences after interpolation
($\{S_n\}$ and $\{T_n\}$).} \label{fig:AR_RD} 
\end{figure} 
The main difference between above two methods (DPCM and DPCMI) is
the number of samples to be coded and the variance of the
respective sequences. The number of samples encoded by DPCM is
half the number encoded by DPCMI, while DPCM encodes a sequence
with 4 times larger variance than that encoded by DPCMI. There is
a clear trade-off between the two methods since reducing the
number of samples will tend to reduce the rate, while increasing
variance will tend to increase the rate per sample.
Fig.~\ref{fig:AR_RD} compares the performance of the two methods
with different AR coefficients. In the figure, the performance of
DPCM is better than that of DPCMI at higher rates but is worse at
lower rates. This shows that at high rates having to encode fewer
samples (DPCM) is better, even though the variance of those
samples is higher. The trade-off is reversed at low bit rates.
Intuitively, DPCM starts to have an error at lower bit-rate than
DPCMI due to the smaller number of samples, but its error is
increased faster than that of DPCMI due to the larger variance.
Therefore at higher rate, the performance of DPCM is better and it
can be worse at lower rate.


Instead of theoretical R-D curves, we now consider operational R-D
curves of general DPCM coders that use a uniform quantizer and an
entropy coder. We assume that a given quantizer has $N$
quantization bins. In the DPCM system, let each bin size be
$\Delta$ (except top and bottom bins assuming that the range of
source is infinite), let the average MSE be $d$ and, after entropy
coding, let the average rate be r. Then, in the DPCMI system, each
bin size can be $\Delta / 2$ and the average MSE is $d/4$ for
$T_n$ since the maximum sample value of $T_n$ is half the maximum
value for $Z_n$. This is because $T_n = (X_n-X_{n-1})/2 = Z_n/2$.
However the number of samples in a given bin is exactly same as
that in a corresponding bin of $Z_n$, so the rate is still $r$
after applying the same entropy coder. Therefore, if the R-D curve
of DPCM passes a point $(r,d)$ then that of DPCMI passes a point
$(2r,d/4)$ (note that $T_n$ and $S_n$ each have rate $r$ in the
DPCMI system). This relation is formulated as
\begin{equation}\label{eq:RD_F}
  G(d) = \frac {1}{2} H(\frac {d}{4}),
\end{equation}
where $G$ and $H$ are the R-D functions of DPCM and DPCMI,
respectively. This also indicates that in general there will be a
trade-off point between the CAI and IAD approaches.

In (\ref{eq:RD_1}), the R-D function is determined by ${Z_n}$
(i.e., the coefficients in a DPCM domain corresponding to a
non-interpolated sequence) and this is not equivalent to the R-D
function of the IAD method (in which the R-D function is
determined by the interpolated sequence after decompression). In
order to evaluate the performance of the IAD method, we first need
to know the R-D performance of the source sequence. The R-D
performance of the difference sequences (generated by open loop
DPCM system) may not be same that of source sequences since the
decoder only has a quantized version of previous sample values.
But, in the orthogonal (or approximately orthogonal) transform
coding case, which is more relevant for image coding (i.e., DCT or
DWT), the distortion in the transform domain is the same as or
close to that in a pixel domain depending on the transform .
Therefore the approximation used in open loop DPCM is not required
and so the analysis approximately holds.
The following shows that in the IAD method, the average MSE can be
decreased after interpolation. Let us assume that the average MSE
of the reconstructed sequence ($\{\hat{X}_n\}$) before
interpolation is $d$ then the sample interpolated between
$\hat{X}_n$ and $\hat{X}_{n+1}$ is $(\hat{X}_n+\hat{X}_{n+1})/2$,
and its average MSE satisfies
\setlength{\arraycolsep}{0.0em}
\begin{eqnarray}\label{eq:MSEI}
E(\frac{\hat{X}_n+\hat{X}_{n+1}}{2}-\frac{X_n+X_{n+1}}{2})^2
&&{}={}E(\frac{(\hat{X}_n-X_n)^2+(\hat{X}_{n+1}-X_{n+1})^2}{4})\nonumber\\
&&~~~~{+}E(\frac{(\hat{X}_n-X_n)\cdot(\hat{X}_{n+1}-X_{n+1})}{2})\nonumber\\
&&{}\leq{} \frac{E(\hat{X}_n-X_n)^2+E(\hat{X}_{n+1}-X_{n+1})^2}{2}\nonumber\\
&&{}={}d~\text{.}
\end{eqnarray}
\setlength{\arraycolsep}{5pt}
In (\ref{eq:MSEI}), inequality comes from
\setlength{\arraycolsep}{0.0em}
\begin{eqnarray}
 &&E((\hat{X}_n-X_n)\cdot(\hat{X}_{n+1}-X_{n+1}))
{}\leq{}
 \frac{E(\hat{X}_n-X_n)^2+E(\hat{X}_{n+1}-X_{n+1})^2}{2}~\text{.}
\end{eqnarray}
\setlength{\arraycolsep}{5pt}
Equality holds only when
$E((\hat{X}_n-X_n)-(\hat{X}_{n+1}-X_{n+1}))^2 = 0$ (i.e., if the
error of two reconstructed pixels is same then the error of the
interpolated pixel is same as the reconstructed pixel). If the
error of $\hat{X}_n$ (i.e., $\hat{X}_n -X_n$) and $\hat{X}_{n+1}$
is uncorrelated and has zero mean then the average MSE of the
interpolated pixel is $d/2$. Therefore average distortion can be
decreased after interpolation. This means interpolated pixels are
not necessarily considered during performance analysis (since
their distortion is always smaller than or equal to that of pixels
coded), and so the analysis based on $R_{1}(D)$ and $R_{2}(D)$ is
still valid for the interpolated sequence. Also, under the
assumption that the R-D functions of DPCM and pixel domains are
similar,
the IAD method provides better performance in a larger
range of different compression ratios.

In a 2-D case, ${X_n}$ and ${Y_n}$ can be considered as
non-interpolated and interpolated images, respectively. Since
transforms, rather than DPCM, are typically used for images, we
can consider ${Z_n}$ and ${T_n}$ (and ${S_n}$) to represent the
coefficients in the transform domain of the non-interpolated and
interpolated images, respectively. Although the order of the
process is same between 1-D and 2-D cases, our analysis of the 1-D
sequence may not be directly applied to a 2-D case  since each
function in the process is not same. But, we can expect that the
IAD method outperforms at very high rates (at least, until the
rates are higher than the rates required for the IAD method
without compression) since the IAD method uses around half the
amount of data. Also, in the IAD method, images may have larger
variance, since it has weaker spatial correlation due to larger
distance between adjacent pixels. So, similar to the result in the
1-D case (shown in Fig.~\ref{fig:AR_RD}), the R-D curve of the IAD
method drops more sharply than that of the CAI method. Therefore,
the R-D curves of IAD and CAI methods may be crossed as bit-rate
is decreased, and depending on the location of the cross point,
the IAD method can provide better performance in practical
applications such as digital cameras.

\section{IMAGE TRANSFORM TO REDUCE REDUNDANCY}
\label{sec:algorithm} 
\begin{figure}[tb]
 \centering
 \includegraphics[width=11cm]{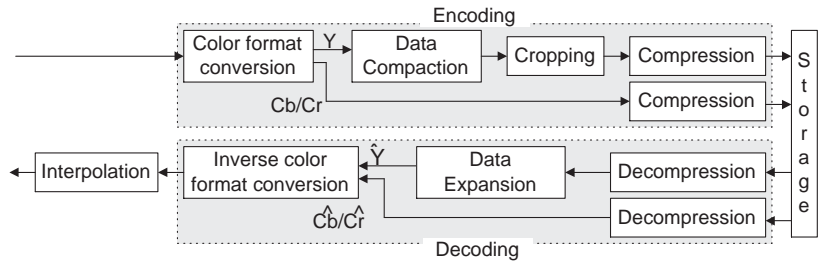}
\caption{Detailed diagram of the encoding and decoding parts of
the proposed IAD method. Luminance (Y) data needs several
transforms due to data location after format conversion, whereas
chrominance (Cb/Cr) data can be coded directly.}
\label{fig:diagnew}
\end{figure}

In order to demonstrate a practical IAD
scheme our first goal is to transform the non-interpolated input
data into a format suitable for general image coders. The input
data of the IAD method consists of only one color value for each
pixel, while in the CAI method there are three color values for
each pixel, obtained by interpolation. In general image coders, it
is assumed that incoming data is uniform (i.e., all pixels have
the same color components) and that the image has rectangular
shape. Our goal is then to design a reversible image transform
that can produce image data suitable for coding (without
increasing the amount of data to be coded). A detailed diagram of
the encoding and decoding blocks of the IAD method is shown in
Fig.~\ref{fig:diagnew}. 
First, we propose a color format conversion algorithm,
since image coders normally use YCbCr format. After format
conversion, luminance (Y) data is not available at every pixel
position. In order to make the Y data compact, we propose a
transform that relocates pixels and removes pixels for which no Y
data is available. Then we show how to encode the resulting data
which no longer has rectangular shape.
\subsection{Color format conversion} \label{sec:format_conversion}
In the CAI method, the data to be compressed is in RGB format
(obtained by interpolating the CFA data). This data is converted
to YCbCr format before compression. In JPEG, normally 4:2:2 or
4:2:0 sampling is used. In JPEG2000, chrominance coefficients in
high frequency bands after wavelet transform are not coded since
the human visual system is less sensitive to the chrominance data.
In the IAD method, to avoid increasing the redundancy, the number
of pixels should not be increased after color format conversion.
While there are several different methods to achieve this, we
choose a method such that 2 green, 1 red and 1 blue pixels are
converted to 2 Y, 1 Cb and 1 Cr pixel values. This is reasonable
since luminance data is more important than chrominance data and
the format conversion can be reversible. We first propose a simple
and fast method based on 2 by 2 blocks and then propose more
complex methods that provide better performance.
\subsubsection{Format conversion based on 2 by 2 blocks} 

\begin{figure}[tb]
\begin{minipage}[b]{1.0\linewidth}
  \centering
  \includegraphics[width=9cm]{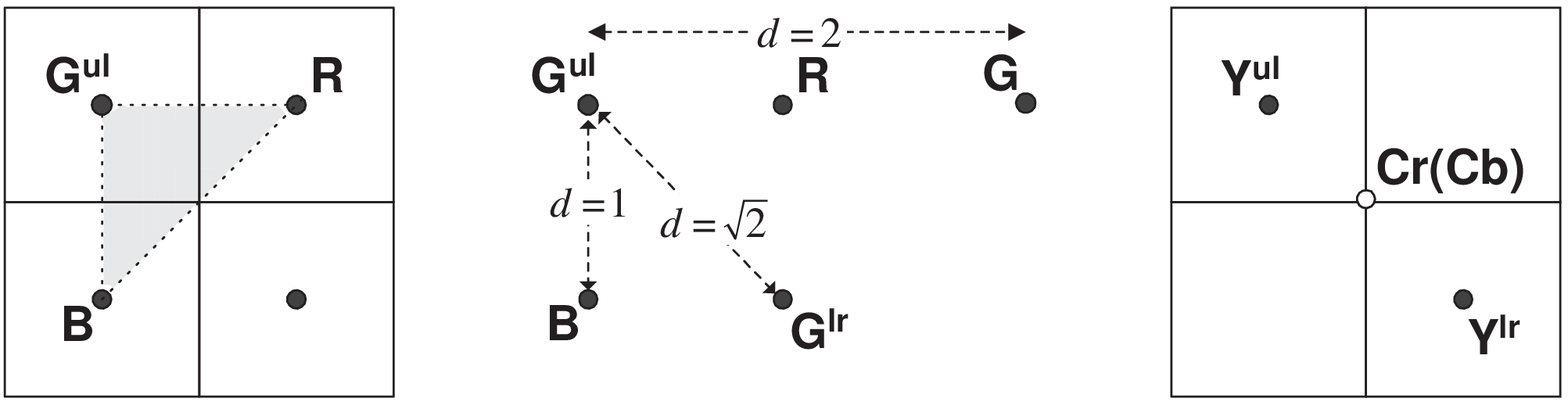}
  \centerline{(a) \hspace{2.7cm}(b)\hspace{2.7cm}(c)}\vspace{-0.0cm}
\end{minipage}
\caption{The gray region in (a) indicates the possible location of
Y data after the format conversion. (b) shows the distance between
two green (or luminance) pixels. (c) shows the location of Y and
Cr (Cb) data in a 2 by 2 block.} \label{fig:distance}
\vspace{0.0cm}
\end{figure}
In this format conversion, each 4 pixel block contains 2
green, 1 red and 1 blue pixels. Then two luminance and two
chrominance values (i.e., Cb and Cr) are obtained by using one
green pixel for each of the two luminances, and using the average
of the two green pixels for the chrominance calculation. This
operation can be represented as follows :
\begin{equation}\label{eq:fc}
\left[\hspace{-0.1cm} \begin{array}{c} Y^{ul} \\ Y^{lr} \\ Cb \\
Cr
\end{array} \hspace{-0.1cm} \right] =
\left[\hspace{-0.1cm} \begin{array}{clcr} a_{11} & a_{12} & 0 & a_{13} \\
a_{11} & 0 & a_{12} & a_{13} \\ a_{21} & \frac{a_{22}}{2} & \frac{a_{22}}{2} & a_{23} \\
a_{31} & \frac{a_{32}}{2} & \frac{a_{32}}{2} & a_{33}
\end{array} \hspace{-0.1cm}\right]\hspace{-0.2cm}
\left[\hspace{-0.1cm} \begin{array}{c} R \\ G^{ul} \\ G^{lr} \\ B
\end{array} \hspace{-0.1cm}\right] + \left[\hspace{-0.1cm} \begin{array}{c} 0 \\ 0 \\ 128 \\ 128
\end{array} \hspace{-0.1cm}\right],
\end{equation}
where, as shown in Fig.~\ref{fig:distance} (b) and (c), the
superscripts $ul$ and $lr$ indicate the upper left and lower right
positions in a 2 by 2 CFA block, respectively. The coefficients
$a_{ij}$ are the (i,j) coefficients of the standard RGB to YCbCr
conversion matrix~\cite{Hamilton92JFIF} defined as follows:
\begin{equation}\label{eq:rgb2ycbcr}
\left[\hspace{-0.1cm} \begin{array}{c} Y \\ Cb \\ Cr
\end{array} \hspace{-0.1cm}\right] =
\left[\hspace{-0.2cm} \begin{array}{clcr} 0.297 & 0.587 & 0.114 \\
-0.169 & -0.331 & 0.500 \\ 0.500 & -0.419 & -0.081
\end{array} \hspace{-0.2cm}\right]\hspace{-0.2cm}
\left[\hspace{-0.1cm} \begin{array}{c} R \\ G \\ B
\end{array} \hspace{-0.1cm}\right] + \left[\hspace{-0.2cm} \begin{array}{c} 0 \\ 128 \\ 128
\end{array} \hspace{-0.2cm}\right]
\end{equation}

We now need to decide what the location of these
$Y^{ul}Y^{lr}CbCr$ pixels should be. For the $Cb$ and $Cr$ data,
each component could be located in any fixed position in the 2 by
2 block, since only one value of each chrominance is generated for
the block. In the Y data case, however, one ($Y^{ul}$) should be
located in the upper left region since $Y^{ul}$ is the weighted
average of $G^{ul}$, $R$ and $B$ (as shown in
Fig.~\ref{fig:distance} (a)) and the other ($Y^{lr}$) should be
located in the lower right region of the block.

In our algorithm, we put the Y data at each green pixel position
because green is roughly 60\% of the Y data (the shape of the Y
image is shown in Fig.~\ref{fig:capt_3} (a)). The location of the
Y data is important, since improperly located Y data induces
artificial high frequency components which can degrade the coding
performance.

This method is simple and fast but YCbCr data of each 2 by 2 block
depends only on the RGB data in that 2 by 2 block. Therefore, the
YCbCr data potentially has more high frequency components than
that generated by using bilinear interpolation (because each block
is treated independently, while in the bilinear interpolation case
each Y term is obtained from a larger set of pixels).
\subsubsection{Format conversion based on larger
blocks} In order to generate smoother YCbCr data, we can consider
a whole image as a block. After generating the RGB data for each
pixel by using bilinear interpolation, Y, Cb and Cr can be
calculated from the RGB data on green, blue and red pixels
respectively. These positions are chosen according to the degree
of influence of each color (i.e., the dominant color components of
Cb and Cr are blue and red, respectively). Although each pixel has
RGB data after interpolation, the amount of YCbCr data is not
increased, since each pixel position has only one component,
either Y, Cb or Cr. This format conversion is also simple but the
reverse format conversion is more complex due to bilinear
interpolation. For example, as in Fig.~\ref{fig:1}, we consider
that the image size is 4 by 4. Then $Y_{22}$ (the luminance value
of the $G_{22}$ position) is calculated as
\begin{equation}\label{eq:YonG}
\begin{split}
Y_{22} &=
\left[0~~\frac{a_{11}}{2}~~0~~0~~\frac{a_{13}}{2}~~a_{12}~~\frac{a_{13}}{2}~~0~~0~\frac{a_{11}}{2}~~0~~0~~0~~0~~0~~0
\right] \\
& \quad \cdot
\left[G_{11}~R_{12}~G_{13}~R_{14}~B_{21}~G_{22}~B_{23}~G_{24}~G_{31}~R_{32}~G_{33}~R_{34}~B_{41}~G_{42}~B_{43}~G_{44}
\right]^T~\text{.}
\end{split}
\end{equation}

But, in the reverse format conversion, $G_{22}$ is calculated as
\begin{equation}\label{eq:GonY}
\begin{split}
G_{22} &= \left[\text{\small{-1243
~-3832~-1079~217~-1763~14895~-1736~
-135~-974~-3658~-868~136~35~-638~26~37}}
\right] \cdot 10^{-4} \\
& \quad \cdot
\left[Y_{11}~Cr_{12}~Y_{13}~Cr_{14}~Cb_{21}~Y_{22}~Cb_{23}~Y_{24}~Y_{31}~Cr_{32}~Y_{33}~Cr_{34}~Cb_{41}~Y_{42}~Cb_{43}~Y_{44}
\right]^T~\text{.}
\end{split}
\end{equation}

%
From the example of (\ref{eq:YonG}) and (\ref{eq:GonY}), we can
see that in general the forward format conversion may be based on
a few neighboring pixels but reverse conversion could require all
pixels in the same block. Therefore, in order to generate the
original RGB data from the YCbCr data, a $w\cdot h$ by $w\cdot h$
reverse format conversion matrix is needed, where $w$ and $h$ are
the width and height of an image respectively.

Although the decoding process (including reverse format
conversion) can be done in a system with high computing power
(e.g., personal computers), the matrix is too large and the
reverse conversion may still be too time consuming. In order to
reduce the computational complexity, the above format conversion
method can be applied to blocks generated by dividing the source
image. Since interpolation is done by using the pixels in the
block, the column (or row) of the reverse format conversion matrix
is reduced to $W\cdot H$, where $W$ and $H$ are the width and
height of a block respectively. In this block based format
conversion, the bilinear interpolation needs to be modified since
only the information of pixels in the same block can be used for
interpolation. For example, the green value on $(i,j)$ position
(which corresponds to a red or blue pixel position) can be
calculated as
\begin{equation}\label{eq:GonR}
G_{ij} = ~\frac{G'_{(i-1)j} + G'_{(i+1)j} + G'_{i(j-1)}+
G'_{i(j+1)} }{I_{(i-1)j}+I_{(i+1)j}+I_{i(j-1)}+I_{i(j+1)}},
\end{equation}
where $G'_{kl} = I_{kl} \cdot G_{kl}$, and $I_{kl}$ is an
indicater function defined as follows.
\begin{equation}\label{eq:I}
I_{kl} = \begin{cases} 1,&~\text{if $kl$ and $ij$ are in the same block} \\
0,&~\text{otherwise.}\\
\end{cases}
\end{equation}
Then $G_{ij}$ is used to calculate $Cb_{ij}$ (or $Cr_{ij})$ at the
blue (or red) position.

Note that, in the decoding process, an approximation method which
uses only neighboring pixels can be used since more distant pixels
do not have a big influence (see the coefficients in
(\ref{eq:GonY})). But in this paper, we focus on a block-based
method in order to avoid the effect of error in format conversion.

The performance comparison among different block size format
conversion is given in
Section~\ref{sec:results_format_conversion}.
\subsection{Nonlinear transform to compact luminance data}
\label{sec:nonlinear} 
\begin{figure}[tb]
\begin{minipage}[b]{1.0\linewidth}
  \centering
  \includegraphics[width=9.5cm]{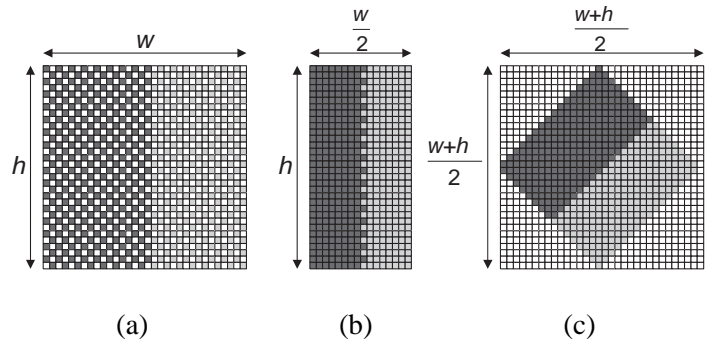} \\
  \centerline{(a) \hspace{2.4cm}(b) \hspace{2.4cm}(c) }\vspace{-0.0cm}
\end{minipage}
\caption{Transform of Y (luminance) image. In the figure, dark and
light gray pixels indicate Y data and white pixels indicate empty
position. (a) indicates quincunx located Y image after format
conversion, (b) and (c) indicate Y image after transform. In (b),
each odd column data is shifted to the left even column and in
(c), each pixel is rotated 45 degree clockwise. }
\label{fig:capt_3}
\end{figure}
After the color format conversion, the Y values are not available
at all the original pixel positions (since the Y data is located
only in the position of the green pixels), so general image
compression methods cannot be directly applied to compress the Y
image. Therefore another reversible transform is needed to change
the Y pixels located on a quincunx lattice (see
Fig.~\ref{fig:capt_3} (a)) to normally located Y pixels (i.e., so
that we obtain a Y image with no blank pixels). As in
Fig.~\ref{fig:capt_3} (b), one possible simple transform is a
horizontal pixel shift where pixels in odd columns are shifted to
the left even column and all odd columns are removed. This
transform can be formulated as
\begin{equation}\label{eq:shift}
if~ x+y = odd,~ \left[ \begin{array}{c} X\\Y
\end{array} \right] =
\begin{cases}
  ~\left[~~ \begin{array}{c} {\frac{x}{2}}\\y
\end{array} ~~\right],~if~ x = even,\\
  ~\left[ \begin{array}{c} {\frac{x-1}{2}}\\y
\end{array} \right],~if~ x = odd,
\end{cases}
\end{equation}
where $(x,y)$ and $(X,Y)$ are the pixel positions in the images
before and after transform, respectively. Here we assume that the
origin is the lower left corner of an image. A vertical shift
transform can be similarly defined, but we focus here on the
horizontal shift transform.

After the transform is performed, a vertical edge (e.g.,
Fig.~\ref{fig:capt_3} (b)) leads to artificial high frequency
components in the vertical direction, which leads to coding
performance degradation. This also happens when the edge is
vertically biased (i.e., steeper than a 45 degree line). Note that
if a vertical shift had been chosen the same problem would arise
with respect to horizontal (and horizontally biased) edges. Thus,
under JPEG coding with a high compression ratio, most of this
artificial high frequency information may be lost.
Spatially weak correlation is another reason that contributes to
making the results worse. If the distance between adjacent pixels
in a row (or a column) in a CFA is assumed to be 1 then, after
horizontal shifting, the vertical and horizontal distances of
adjacent pixels in the Y data are $\sqrt{2}$ and $2$ respectively
(see Fig.~\ref{fig:distance} (b)).

An alternative simple transform to remove blank pixels among Y
data, which does not pose these problems, is a 45 degree rotation
formulated as
\begin{equation}\label{eq:rotation}
\begin{split}
\left[ \begin{array}{c} X \\ Y \end{array} \right] = \frac{1}{2}
\left(  \left[ \begin{array}{clcr} 1 & 1 \\ -1 & 1 \end{array}
\right] \left[ \begin{array}{c} x\\y \end{array} \right] + \left[
\begin{array}{c} -1 \\ w-1 \end{array} \right] \right),\\
~~\text{for }x + y = odd,
\end{split}
\end{equation}
where $w$ indicates the image width. As shown in
Fig.~\ref{fig:capt_3} (c), after rotation, Y data is concentrated
on the center of an image with an oblique rectangular shape. This
transform does not induce artificial high frequencies and the
distances between adjacent pixels in a row or column are now
$\sqrt{2}$. But since the data no longer has a standard
rectangular shape area, some redundancy is added when the boundary
pixels are coded. This is addressed in the next section and the
performance comparison between shift and rotation transform is
presented in Section~\ref{sec:results_nonlinear_transform}.

As shown in (\ref{eq:shift}) and (\ref{eq:rotation}), the
complexity of both methods is low. The rotation method needs 1
comparison, 2.5 addition and 1 shift operation per pixel whereas
the shift method needs 1.5 comparison, 1.25 addition and 0.5 shift
operation.

\subsection{Data cropping for images obtained by the rotation transform}
\label{sec:cropping} After the horizontal shift transform (see
Fig.~\ref{fig:capt_3} (b)) Y data can be directly encoded. But the
shape of Y data after the rotation transform is not rectangular
(see Fig.~\ref{fig:capt_3} (c)) and thus coding the shape bounding
box (i.e., the whole rectangular region that includes the oblique
rectangular shape of Y data) would result in some inefficiency in
the coding. Therefore a proper cropping method is needed to remove
the data outside of the oblique rectangular area containing Y
data.
\subsubsection{Data cropping for JPEG (DCT based coders)}
In JPEG, the size of a DCT block is 8 by 8 and blocks that consist
of blank pixels only (blank blocks) do not need to be coded. In
addition, we do not need to send any side information about the
location of Y data since it can be calculated at the decoder given
the size of the original image. As shown in Fig.~\ref{fig:capt_3}
(c), the number of blank blocks depends on the width and height of
the image and 6 bits (2 bits for a zero DC value and 4 bits for
EOB (end-of-block)) are needed to code a blank block when standard
Huffman tables of JPEG are employed. In case of 512 by 512 images,
out of 4096 blocks, the number of blank blocks is 1984 and without
coding blank blocks, we can save 1488 bytes. The blocks containing
boundary pixels of Y data (boundary blocks) also contain blank
pixels since Y data are in an oblique rectangular shape. As a
result, compared to the shift method, the number of blocks to be
coded is increased by $(w+h)/16$ in case that the width and height
are multiples of 16.

Proper padding methods are needed for boundary blocks since the
discontinuity between blank and data pixels in the block creates
artificial edges that require a significant coding rate. Because
boundary blocks have Y data only in the position of an upper or
lower triangular region, padding can be simply done by diagonal
mirroring using a data copy where the source data position is
determined by table look-up. Four different look-up tables are
needed since each boundary needs a different pattern of a data
copy. Better performance can be achieved by using low-pass
extrapolation (LPE)~\cite{Kaup99CSVT} or shape adaptive DCT
(SA-DCT)~\cite{Sikora95SP}\cite{Kaup99CSVT}\cite{Stas99CSVT}. LPE
is relatively simple and provides good R-D performance whereas
SA-DCT provides better performance but is more complex.

\subsubsection{Data cropping for SPIHT (DWT based coders)}
Contrary to JPEG, SPIHT is not a block based coder and so the
method used with JPEG cannot be applied. Therefore we need to
introduce a new coding method in order to code Y data in the
oblique rectangular area only. In the still image coding of
MPEG-4, arbitrarily shaped objects are coded by using shape
adaptive DWT (SA-DWT)~\cite{Li00CSVT}. SA-DWT uses a length
adaptive 1-D DWT after finding the first non-blank pixel in the
line and the low-pass (high-pass) wavelet coefficients are placed
into the corresponding location in the low-pass (high-pass) band
(i.e., the shape is reserved in each band after transform). One of
the good features of SA-DWT is that the number of coefficients
after SA-DWT is identical to the number of data pixels. In order
to code data pixels only, we employ SPIHT with SA-DWT. But without
modifying entropy coding in SPIHT, some redundancy is still added
since SPIHT uses a two by two block arithmetic coding algorithm.
\begin{figure}[tb]
\centering
\includegraphics[width=8.5cm]{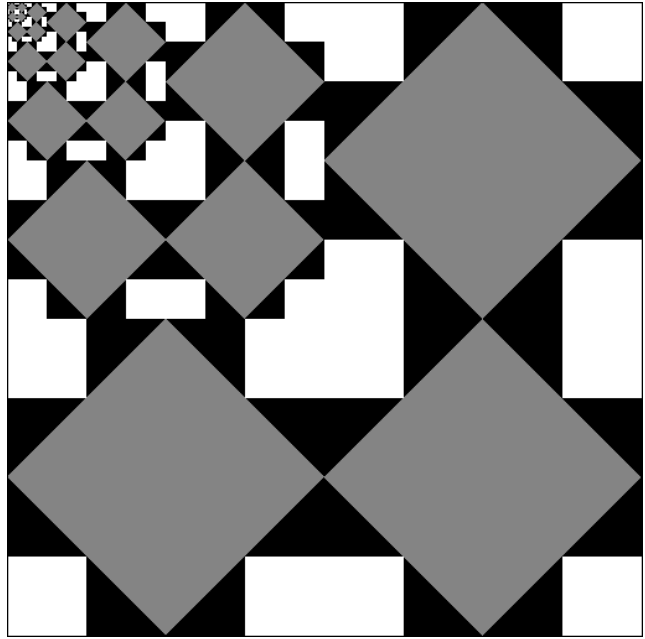}
\caption{The coefficients map after SA-DWT (when the width and
height of images are 512). Gray regions indicate meaningful
coefficients and black and white regions indicate blank
coefficients.} \label{fig:mask}
\end{figure}
In Fig.~\ref{fig:mask}, only the gray regions contain meaningful
coefficients after SA-DWT. Out of 16 two by two blocks in the
lowest frequency band, only 4 blocks located in each corner
consist of blank coefficients. Since all descendants of these
blocks (white regions in the figure) are blank coefficients, these
regions are not coded. But blank coefficients in black regions in
the figure are involved in coding due to the entropy coding scheme
of SPIHT, and since they are not skipped some redundancy is
introduced.

The complexity of the SA-DWT is no more than that of conventional
DWT for the shape bounding box size image~\cite{Li00CSVT}. Here,
the width and height of the shape bounding box are $(w+h)/2$, but
the complexity is much lower since the shape is at most half the
size of the shape bounding box. Also, a simpler SA-DWT method can
be applied since the shape is convex. In fact, after finding the
non-blank data position (which can be calculated from the width
and height of the image), the complexity is about half of DWT for
interpolated images. Also, in the entropy coding of SPIHT, the
white region in Fig.~\ref{fig:mask} is not coded and the tree
related to the black region is terminated early when all the
descendent are blank coefficients.
%
\subsection{Influence of chrominance data over luminance data }
In the IAD algorithms, one Cb (Cr) data is chosen out of 4 CFA
pixels and the width and height of the Cb (Cr) image are $w/2$ and
$h/2$, respectively. Therefore the data size is reduced to a
quarter of that in the conventional technique whereas the pixel
distance is doubled.
%
But contrary to the CAI method, in which the coding results of
luminance and chrominance data are fully separated, chrominance
data with large distortion can add distortion to luminance data
after interpolation, and vice versa.

For the case of format conversion with 2 by 2 blocks, the coding
error in RGB data is calculated as follows.
\begin{equation}\label{eq:error_influence}
\left[\hspace{-0.1cm} \begin{array}{c} e(R) \\ e(G^{ul}) \\ e(G^{lr}) \\
e(B)
\end{array} \hspace{-0.1cm}\right] =
\left[\hspace{-0.1cm} \begin{array}{clcr} a_{11} & a_{12} & 0 & a_{13} \\
a_{11} & 0 & a_{12} & a_{13} \\ a_{21} & \frac{a_{22}}{2} & \frac{a_{22}}{2} & a_{23} \\
a_{31} & \frac{a_{32}}{2} & \frac{a_{32}}{2} & a_{33}
\end{array} \hspace{-0.1cm}\right]^{-1}
\left[ \hspace{-0.1cm}\begin{array}{c} e(Y^{ul}) \\ e(Y^{lr}) \\
e(Cb) \\ e(Cr)
\end{array} \hspace{-0.1cm}\right],
\end{equation}
where $e(\cdot)$ is the error of each component due to lossy
coding. Since the final Y data (after interpolation) is calculated
from the distorted RGB data, i.e., from distorted YCbCr data, the
error in the final Y data depends on the quantization errors in
both Y and Cb (Cr) data. For example, after applying bilinear
interpolation, the error of the final Y on $G_{33}$ (in
Fig.~\ref{fig:1}) is calculated as follows.
\begin{equation}\label{eq:error_influence_ex}
\begin{split}
e(\tilde{Y_{33}}) &=
\frac{a_{11}}{2}(e(R_{32})+e(R_{34}))+a_{12}~e(G_{33})+\frac{a_{13}}{2}(e(B_{23})+e(B_{43}))\\
&=0.897~e(Y_{33})+0.075~(e(Y_{31})+e(Y_{42}))+0.029~(e(Y_{13})+e(Y_{24}))-0.103~e(Y_{44})\\
&~~+0.210~(e(Cr_{32})-e(Cr_{34}))+0.101~(e(Cb_{43})-e(Cb_{23})),
\end{split}
\end{equation}
where $\tilde{Y}$ is the final Y, and this comes from
(\ref{eq:YonG}) and (\ref{eq:error_influence}). In
(\ref{eq:error_influence_ex}), the error in the final Y depends on
not only the error in Y but also the error difference between two
Cb (Cr) involved in the interpolation. Therefore, in order to
maximize the quality of final Y and Cb (Cr) data under given bit
budget, bit allocation between Y and Cb (Cr) data needs to be
considered.

In SPIHT, each component is coded separately and there are no
explicit mechanisms for bit allocation, whereas in JPEG bit
allocation to each component cannot be explicitly controlled and
is determined by the chosen quantization tables and the data
characteristics. Therefore in SPIHT, it is necessary to determine
the bit allocation between luminance and chrominance data based on
human visual sensitivity to each component. Moreover, in the
proposed methods, the bit-rate of one component affects the
quality of the other components, so the overall performance
changes depending on the bit allocation.

\begin{figure}[tb]
\centering
 \begin{tabular}{cc}
 \includegraphics[width=7.6cm]{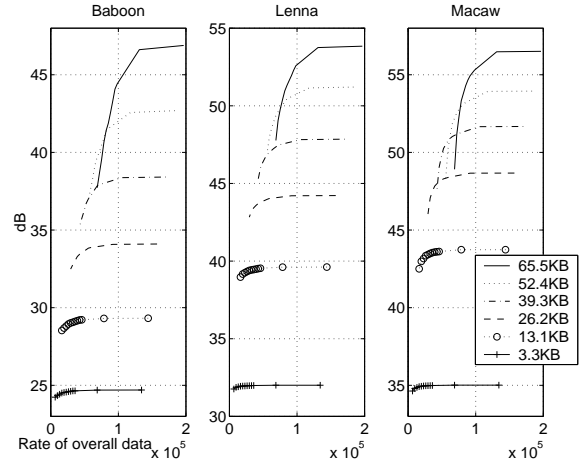} \\
   (a)\\
 \includegraphics[width=7.9cm]{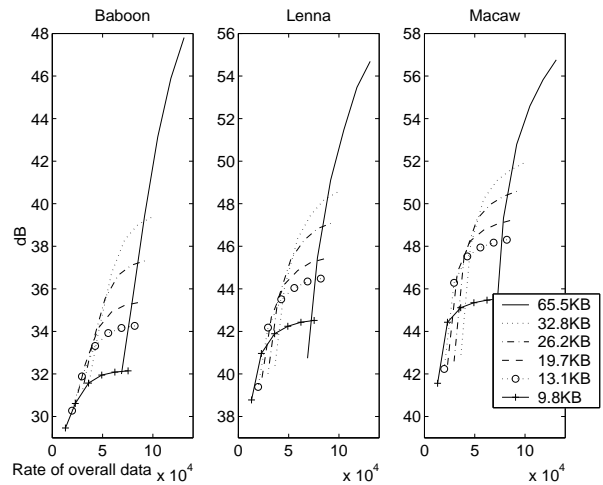} \\
   (b)\\
  \end{tabular}
\caption{The curves indicate (a) luminance and (b) chrominance
PSNR after interpolation depending on the overall bit-rate. SPIHT
is used as a compression method and PSNR is calculated from the
distortion between the interpolated image before compression and
the final output image of proposed methods. The bit-rates shown in
the box correspond to (a) luminance and (b) chrominance data.}
\label{fig:bit_allocation}
\end{figure}

Here, we simply consider bit allocation based on the quality of
luminance data since the human visual system is more sensitive to
luminance data. Fig.~\ref{fig:bit_allocation} (a) shows the
quality change of luminance data after interpolation depending on
the overall bit-rate and the bit-rate of the luminance data.
The intersection of the curves in the figure shows more bit budget
for Y does not guarantee higher PSNR of Y. Also, we can find that
the PSNR of Y is maximized when the bit budget of Y is roughly
$80\%$ of the overall bit budget. Similarly,
Fig.~\ref{fig:bit_allocation} (b) shows the quality change of
chrominance data after interpolation. Although the PSNR decreases
sharply, this happens in the range where the bit-rate of luminance
data is lower than that of chrominance data. In general, the
bit-rate of luminance data is higher than that of chrominance data
and this drop does not have a significant effect. This also
guarantees that we can focus on the quality of luminance data.

Since the R-D characteristics are different for each image, we
fixed the bit-rate of Cb (Cr) data to be a quarter of that of Y
data (i.e., $66.7\%$ of the overall budget). 

\section{Experimental results and discussion} \label{sec:capt_results}
In order to confirm the validity of the IAD algorithms, we
implemented these algorithms (horizontal shift transform with 2 by
2 block format conversion and rotation transform with 2 by 2 and
64 by 64 block format conversion) and compared the results with
those obtained with CAI (JPEG with 4:2:2 format and SPIHT)
methods. Due to the lack of CFA raw data, we generate CFA raw data
by using test images such as ``Baboon'', ``Lenna'' and ``Macaw''
(H : 512, W : 512, 24 bit color, 786.432KB). In fact, what we
obtain is not CFA raw data, since in these images all image
processing functions, except interpolation, have been already
done. Since our main focus has been on interpolation and
compression methods, other image processing parts are not
considered. Our results are the same as the results achieved when
all image processing functions are done before compression and
interpolation. As we mentioned in Section~\ref{sec:toy}, in order
to compare the performance,
we consider interpolated images without
compression as the reference images.

The results of the proposed and conventional algorithms are
compared using the PSNR of luminance (and chrominance) data and
the average $\Delta{E}$ in CIELAB color space at each target
bit-rate. Parts of experimental results are also shown in a
webpage~\cite{Lee04Data}. In the webpage, visual comparison is
provided under a fixed compression ratio ($15:1$) and the
difference of bit-rate under near-lossless coding is provided by
using fixed PSNR (48dB). The compression ratio used in digital
cameras is about $3:1$ to $20:1$ depending on vendors and user
setting~\cite{HP}\cite{KODAK}, and the visual difference of the
results is not clearly noticeable although the IAD method provides
large PSNR gain in this low compression range. But with the IAD
method, same quality can be achieved with much lower bit-rate.

We first compare the performance of different color format
conversion and nonlinear transform with cropping proposed in
Sections~\ref{sec:format_conversion} and~\ref{sec:nonlinear}
(\ref{sec:cropping}), respectively. After that we compare the
overall performance of the IAD method to that of the CAI method.
\subsection{Color format conversion} \label{sec:results_format_conversion}
\begin{figure}[tb]
\begin{minipage}[b]{1.0\linewidth}
\centering
\includegraphics[width=8.8cm]{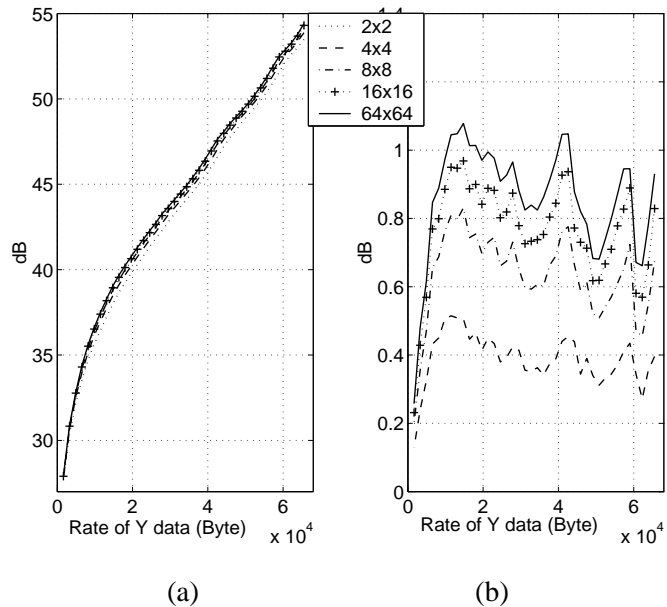}
\centerline{(a) \hspace{3.4cm} (b)}
\end{minipage}
\caption{(a) Coding performance comparison of Lenna image using
different color format conversion methods. (b) Coding gain of the
format conversion using larger blocks against the format
conversion with 2 by 2 blocks. Luminance data are coded by using
SPIHT with shape adaptive DWT (SA-DWT) after rotation transform
and the PSNR is calculated with $Y$ and $\hat{Y}$ in
Fig.~\ref{fig:diagnew} .} \label{fig:block}
\end{figure} 
The coding performance of the format conversion with different
block sizes is shown in Fig.~\ref{fig:block}. Since the
interpolated data at boundary pixels of each block is less smooth,
the interpolation method with a smaller number of boundary pixels
can give a better result. For example, 75\% of Y data are on block
boundaries when 4 by 4 blocks are used whereas 12.3\% are on block
boundaries when 64 by 64 blocks are used. Therefore as shown in
Fig.~\ref{fig:block}, the format conversion with larger blocks
gives better results than that with smaller blocks although the
complexity of decoding is higher.
\subsection{Nonlinear transform with cropping} \label{sec:results_nonlinear_transform}
\begin{figure}[tb]
\begin{minipage}[b]{1.0\linewidth}
\centering
\includegraphics[width=8.6cm]{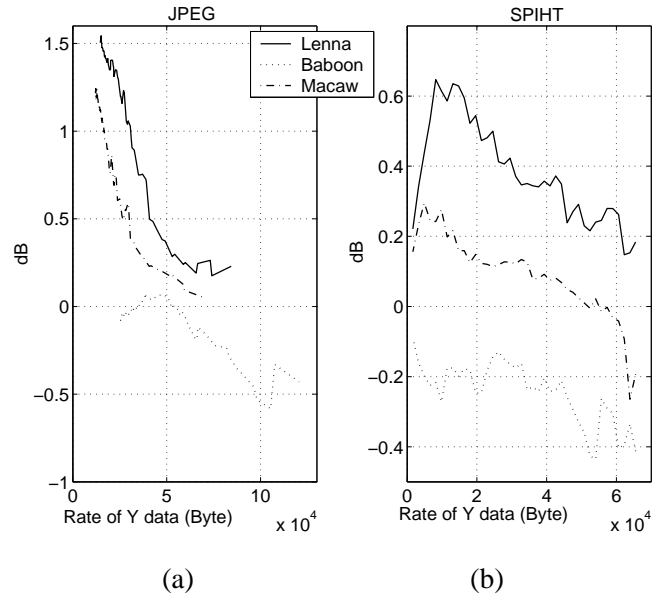}
\centerline{(a) \hspace{3.4cm} (b)}
\end{minipage}
\caption{Luminance PSNR difference between the rotation and
horizontal shift methods after compression by using (a) JPEG and
(b) SPIHT. 2 by 2 block format conversion is used in both cases.}
\label{fig:tran_diff}
\end{figure}
The performance of horizontal shift and rotation methods after
coding is shown in Fig.~\ref{fig:tran_diff}. Since, in JPEG coding
(shown in (a)), high frequency components introduce more errors
due to larger quantization step sizes chosen for them, the
horizontal shift method, which introduces more high frequency
components, results in worse performance. Also in JPEG coding, an
image is efficiently coded by using EOB, so that increased high
frequency energy translates into more bits, as the EOB happens
later on average in a zigzag scan. Thus, as we expected, the
horizontal shift transform generates more high frequency
components and gives a worse result. But for the ``Baboon'' image,
the image itself contains large high frequency components and most
coefficients cannot be coded as EOB, therefore the high frequency
components induced by the shift method have less of an impact than
for other images. In this case, the added redundancy coming from
the data shape of the rotation method may lead to worse
performance than the horizontal shifting, given that the effect of
additional high frequencies is not as significant for these
images.

Contrary to JPEG coding, SPIHT coding does not compress high
frequency components with larger quantization values. Instead it
uses bit-plane encoding at all frequencies and the number of
bit-planes transmitted at a given bit-rate is roughly the same at
all frequencies. Therefore the coding performance of the shift
method is comparable to that of the rotation method. But if the
source is simple (i.e., not having large energy in high frequency
bands) then the shift method provides worse energy compaction and
so the result is worse (as shown in the case of ``Lenna" image).

Figs.~\ref{fig:tran_diff} (a) and (b) show that the coding gain of
the rotation method is decreased as the bit-rate is increased,
except when the bit-rate is less than 10KB. In the low bit-rate
region, small coefficients are quantized to zero, therefore most
coefficients are not transmitted because an EOB has been reached
(in JPEG coding) or only a small number of coefficients is coded
(in SPIHT coding). But in the shift method, many coefficients are
large and cannot be quantized to zero. Therefore higher coding
gain is achieved with the rotation method. As quantization values
become smaller, the coefficients of the rotation method (which are
quantized to zero in a low bit-rate region) are no longer
quantized to zero and the bit-rate increases sharply. Instead,
most coefficients of the shift method are already non-zero (in the
low bit-rate region) and the bit-rate is increased more slowly.
Therefore the coding gain of the rotation method is reduced as the
bit-rate becomes higher.
\subsection{Overall performance}
\begin{figure*}[tb]
\centering
 \begin{tabular}{cc}
   \includegraphics[width=7.8cm]{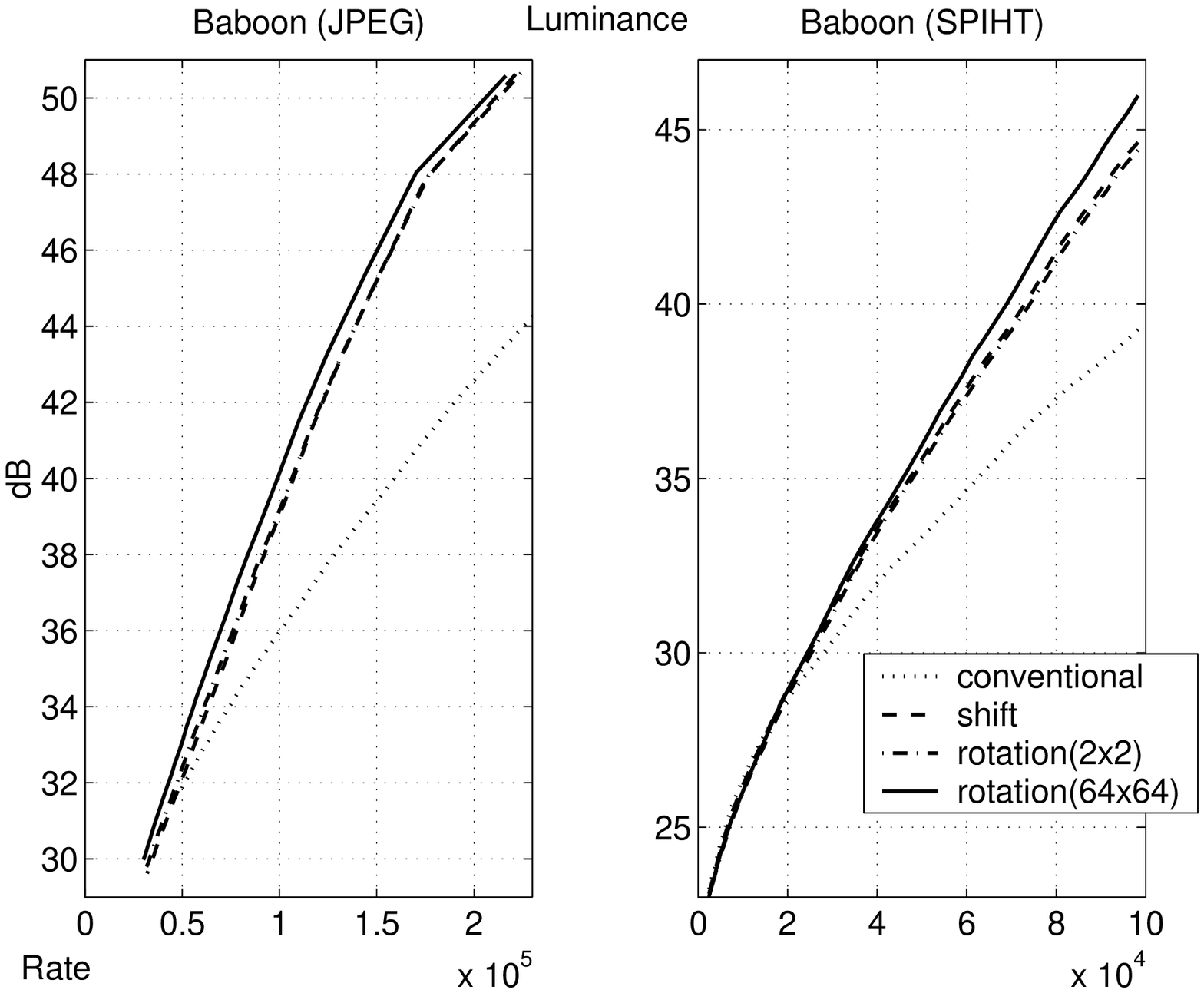} &
   \includegraphics[width=7.8cm]{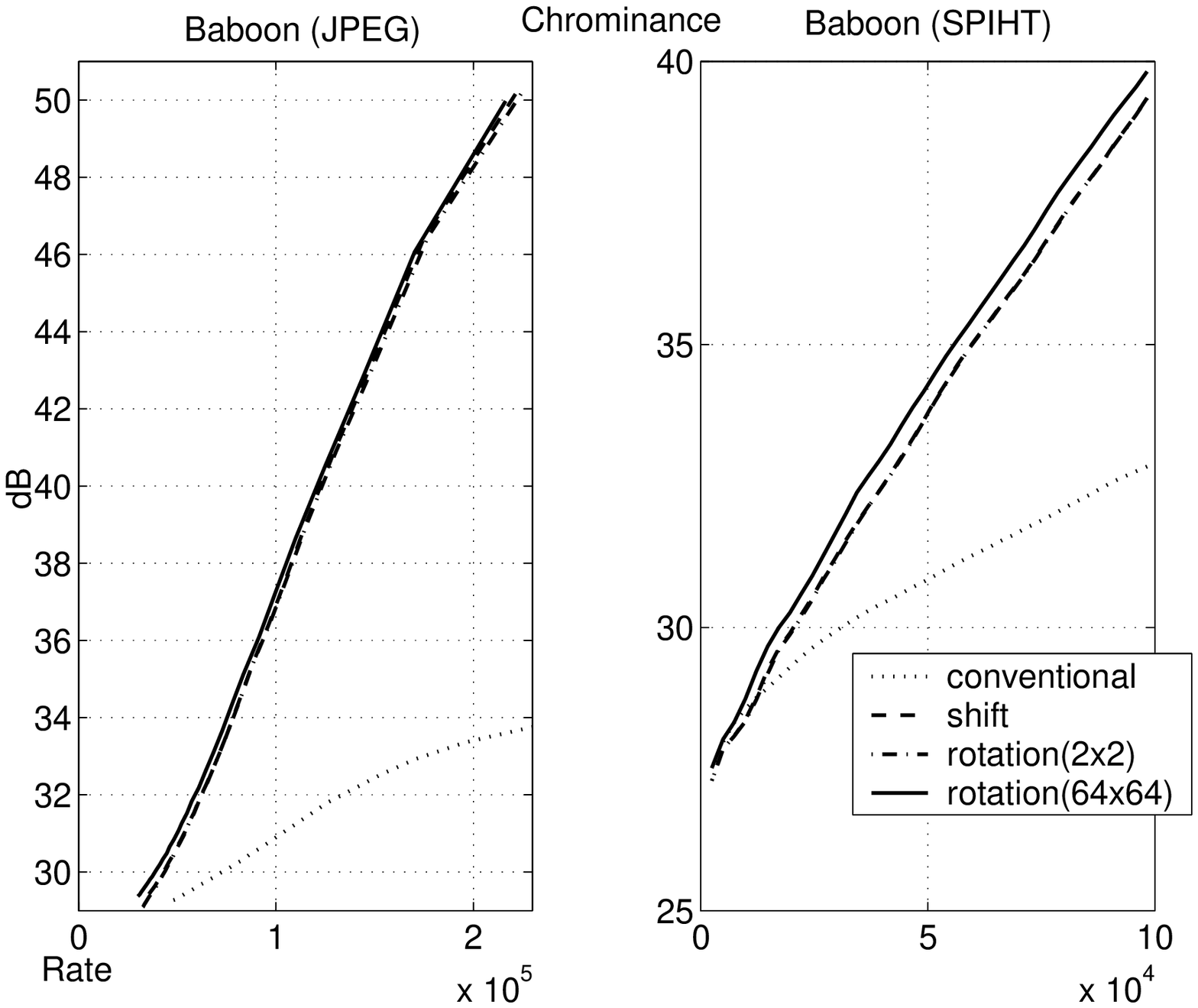} \vspace{-0.3cm}\\
   (a)&(b)\\
   \includegraphics[width=7.8cm]{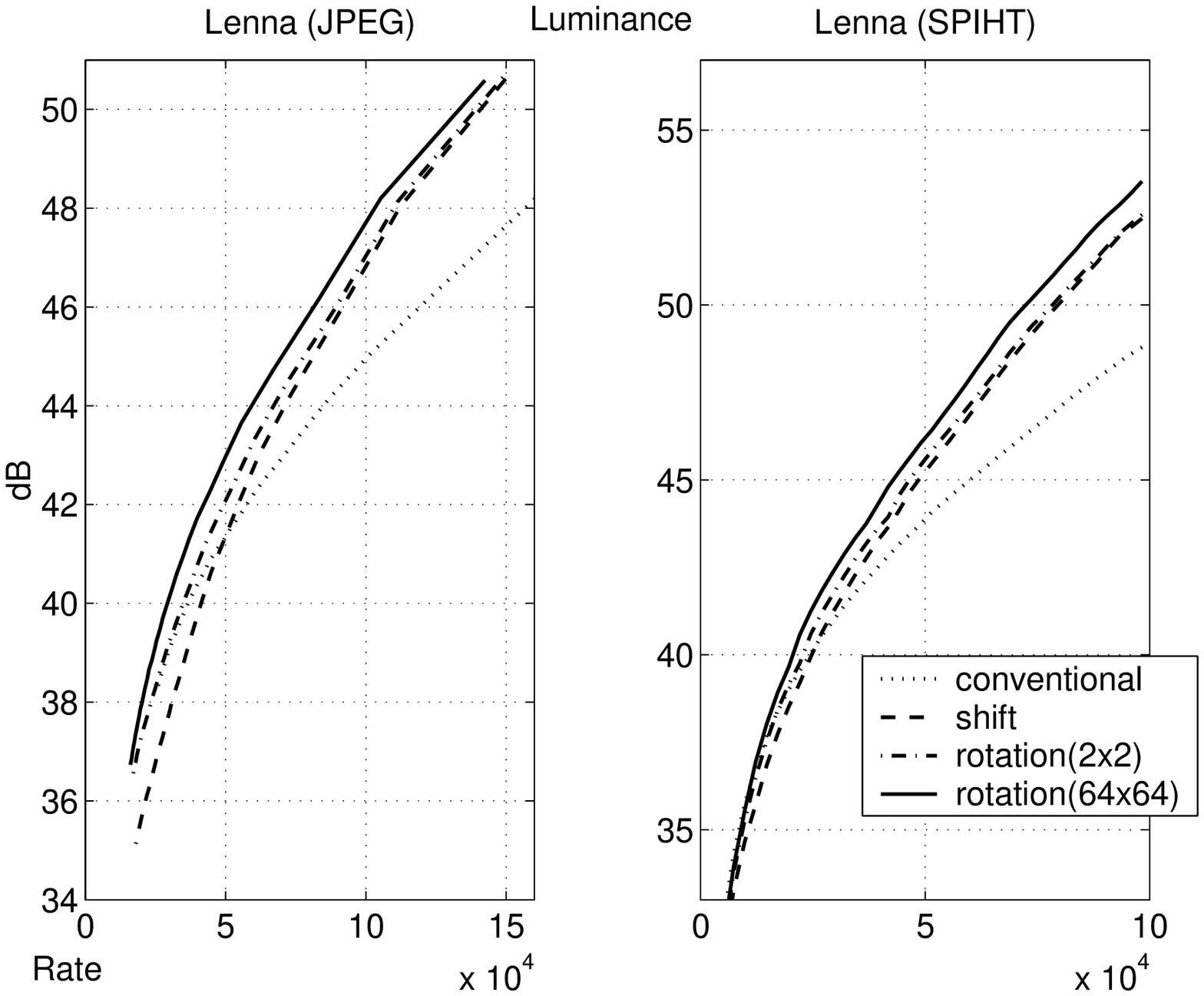} &
   \includegraphics[width=7.8cm]{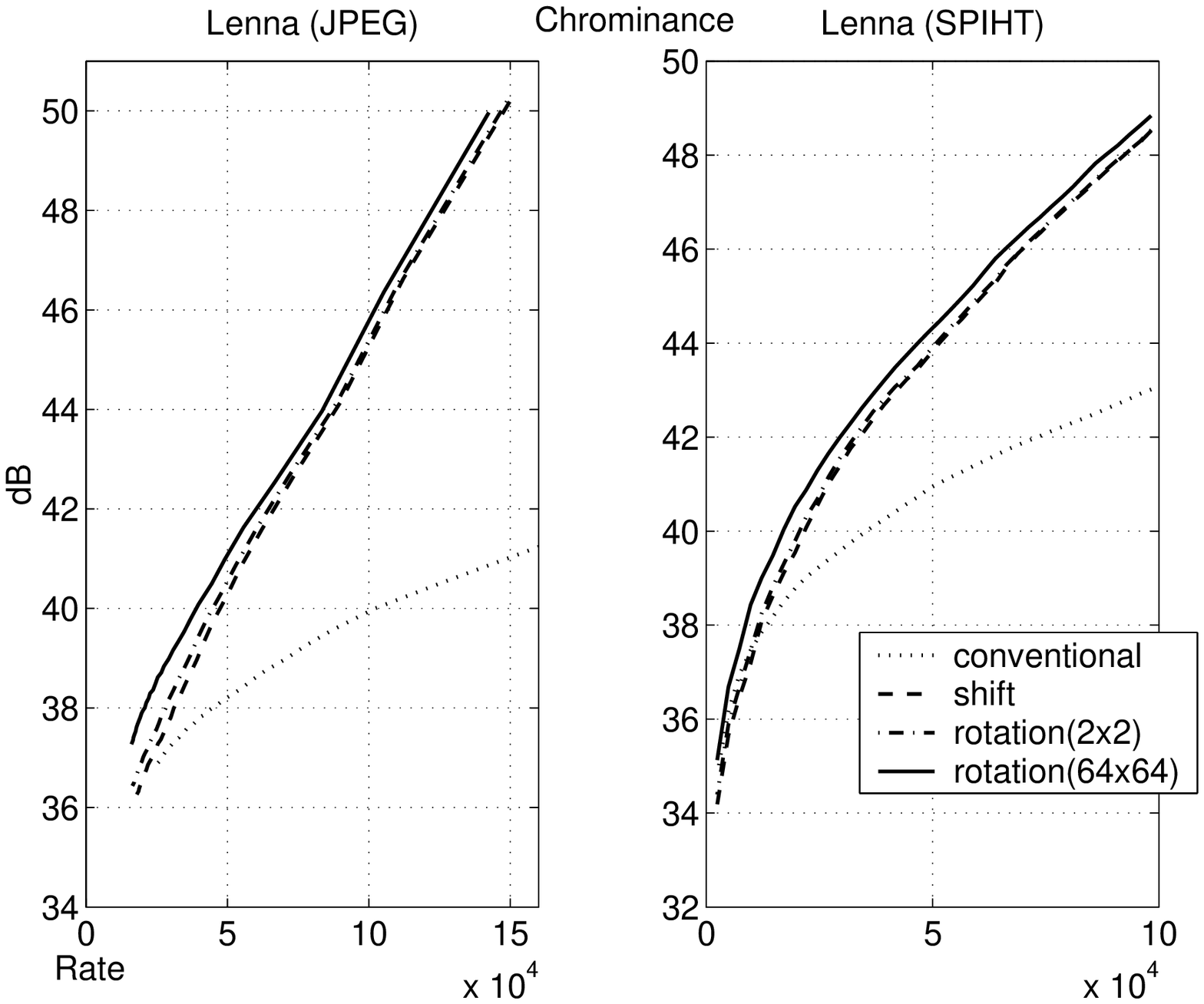}\vspace{-0.3cm} \\
   (c)&(d)\\
   \includegraphics[width=7.8cm]{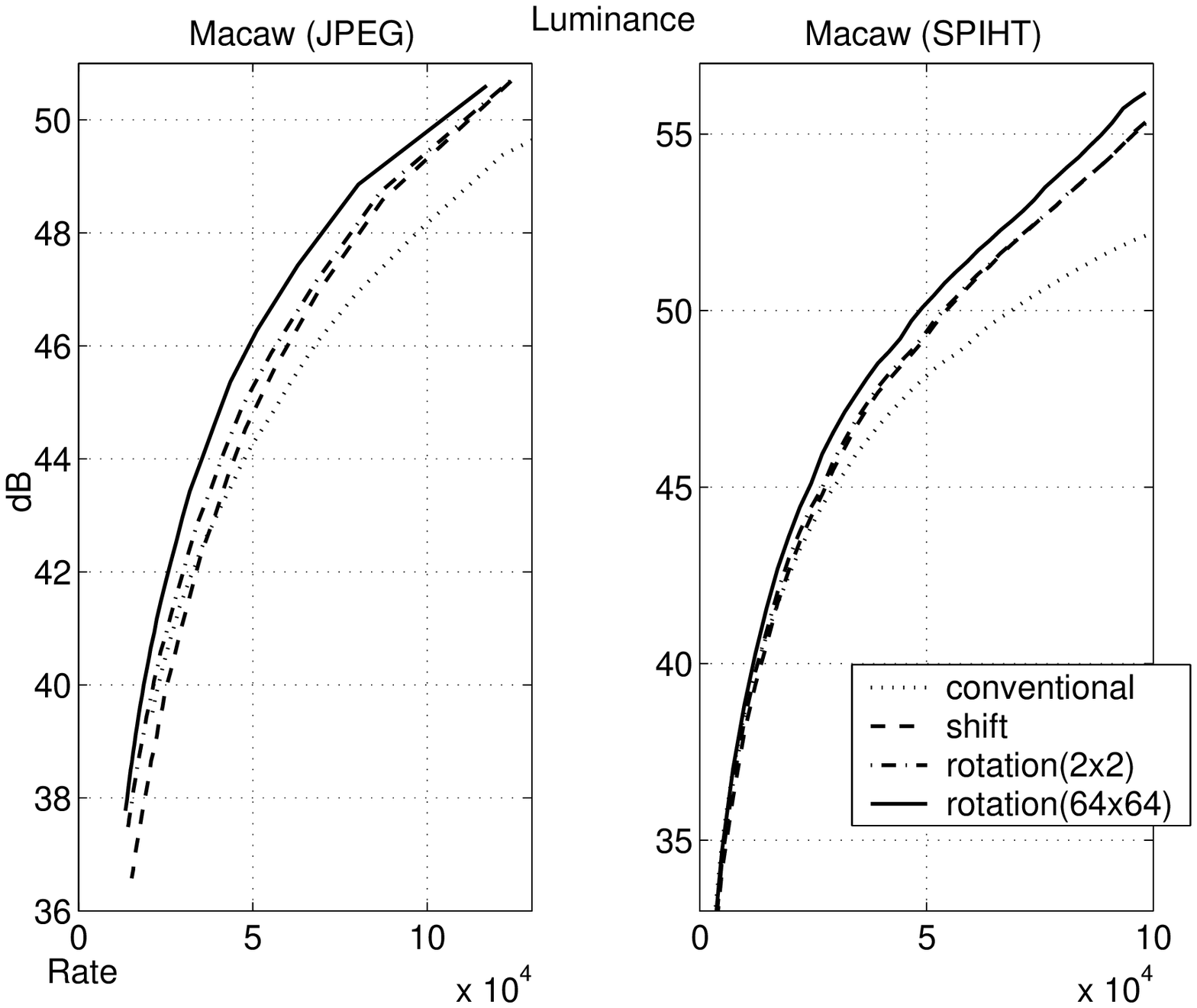} &
   \includegraphics[width=7.8cm]{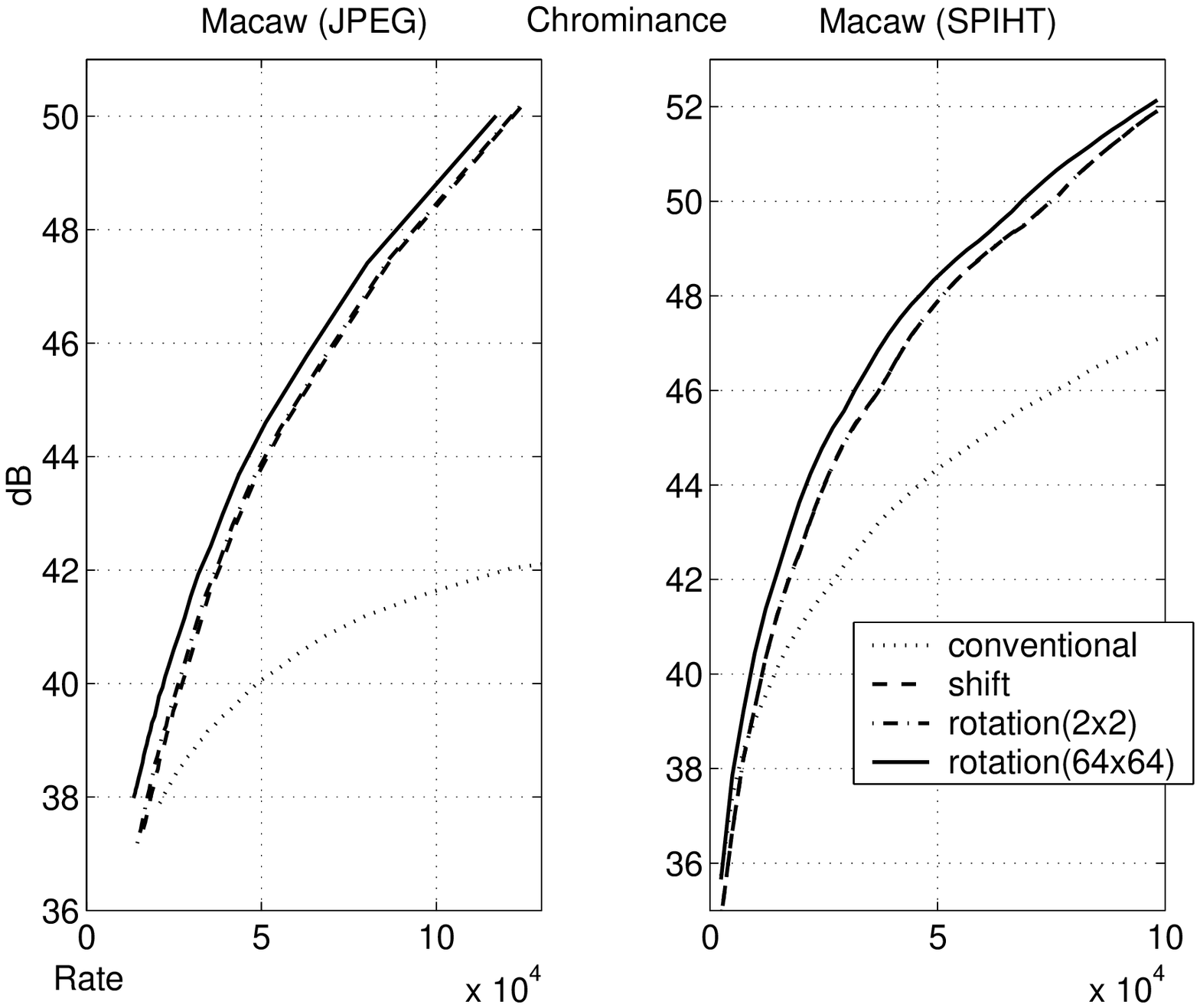}  \vspace{-0.3cm} \\
   (e)&(f)\\
  \end{tabular}
\caption{The curves indicate the luminance and chrominance PSNR
after applying overall coding schemes. \textit{shift} and
\textit{rotation} (2x2 and 64x64) indicate non-linear transform
used in the IAD method and \textit{conventional} indicates the CAI
method.} \label{fig:final_result}
\end{figure*}

\begin{figure}[tb]
\centering
   \includegraphics[width=8.8cm]{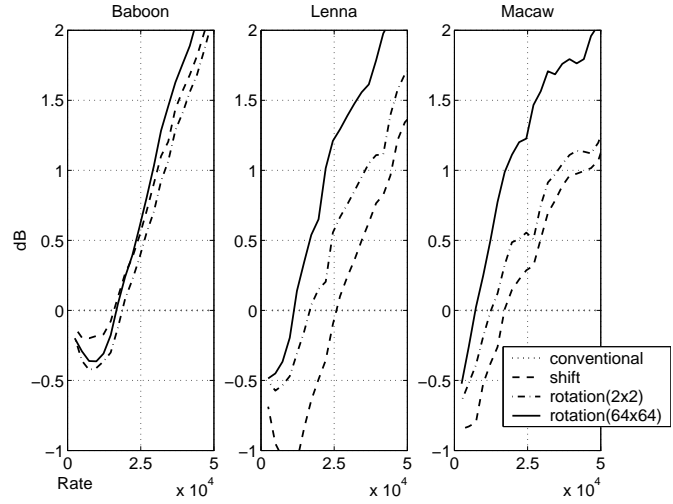}
\caption{The PSNR gain of different proposed methods against the
conventional method. Vertical and horizontal axes indicate the
luminance PSNR gain and overall bit-rate respectively and SPIHT is
used as a compression method. }
\label{fig:diff_l}
\end{figure}

As shown in Figs.~\ref{fig:final_result} (a),(c) and (e), the IAD
algorithms achieve better luminance PSNR except for low bit-rates,
depending on format conversion and compression methods. With JPEG
compression, the PSNR of the shift method drops sharply and the
performance of this method is worse than that of CAI methods in
case that the bit-rate is approximately under 50KB (i.e., the
compression ratio is roughly $15:1$) whereas the rotation methods
outperform the CAI method under all other compression ratios used
in~\cite{JPEGSW}. With SPIHT compression, the performance of shift
and rotation with 2 by 2 block format conversion methods is
similar (see Fig.~\ref{fig:tran_diff} (b)) and they outperform the
CAI method when the bit-rate is over 20KB or 25KB (i.e., a
compression ratio is $39:1$ or $31:1$) as shown in
Fig.~\ref{fig:diff_l}. As expected, the rotation with 64 by 64
block transform method gives a better result than other proposed
methods. Under same bit-rate, IAD algorithms can assign more bits
for each pixel since the IAD algorithms only use approximately
half of luminance data. This is the reason why the IAD methods
outperform the CAI method. Also, as shown in our analysis (see
Fig.~\ref{fig:AR_RD}), IAD algorithms outperform in a wider range
of PSNR with the ``Baboon'' image (low spatial correlation) than
with the ``Lenna'' and ``Macaw'' images (high spatial
correlation).

In the chrominance data cases as in Figs.~\ref{fig:final_result}
(b),~(d) and (f), the PSNR gain is even higher (though PSNR is not
so meaningful in color components). In the CAI algorithm, if the
4:2:2 format is used for JPEG compression then two adjacent pixels
use same color data and some color information is lost. But in the
IAD algorithm, color format conversion is reversible and all color
information can be presented. Although, even in the CAI method,
there is no color information loss if JPEG with 4:4:4 format is
used, the bit-rate for the color information is increased and so,
by using this increased bit-rate, lower compression ratio can be
applied in the IAD algorithm. In our experiments, chrominance data
compression with 4:4:4 format
 is tested with SPIHT compression. Since the size of chrominance data
of the CAI algorithm is 4 times larger than that of proposed ones,
the bit budget per pixel of IAD algorithms is 4 times larger than
that of conventional one and this gives a large PSNR gain.

Although shift and rotation with 2 by 2 block transform provide
exactly the same chrominance data (since both transforms use the
same 2 by 2 color format conversion), the chrominance PSNRs of the
two algorithms are not identical. This shows that the luminance
distortion also affects the quality of chrominance data.

\begin{figure*}[tb]
\centering
\includegraphics[width=15cm]{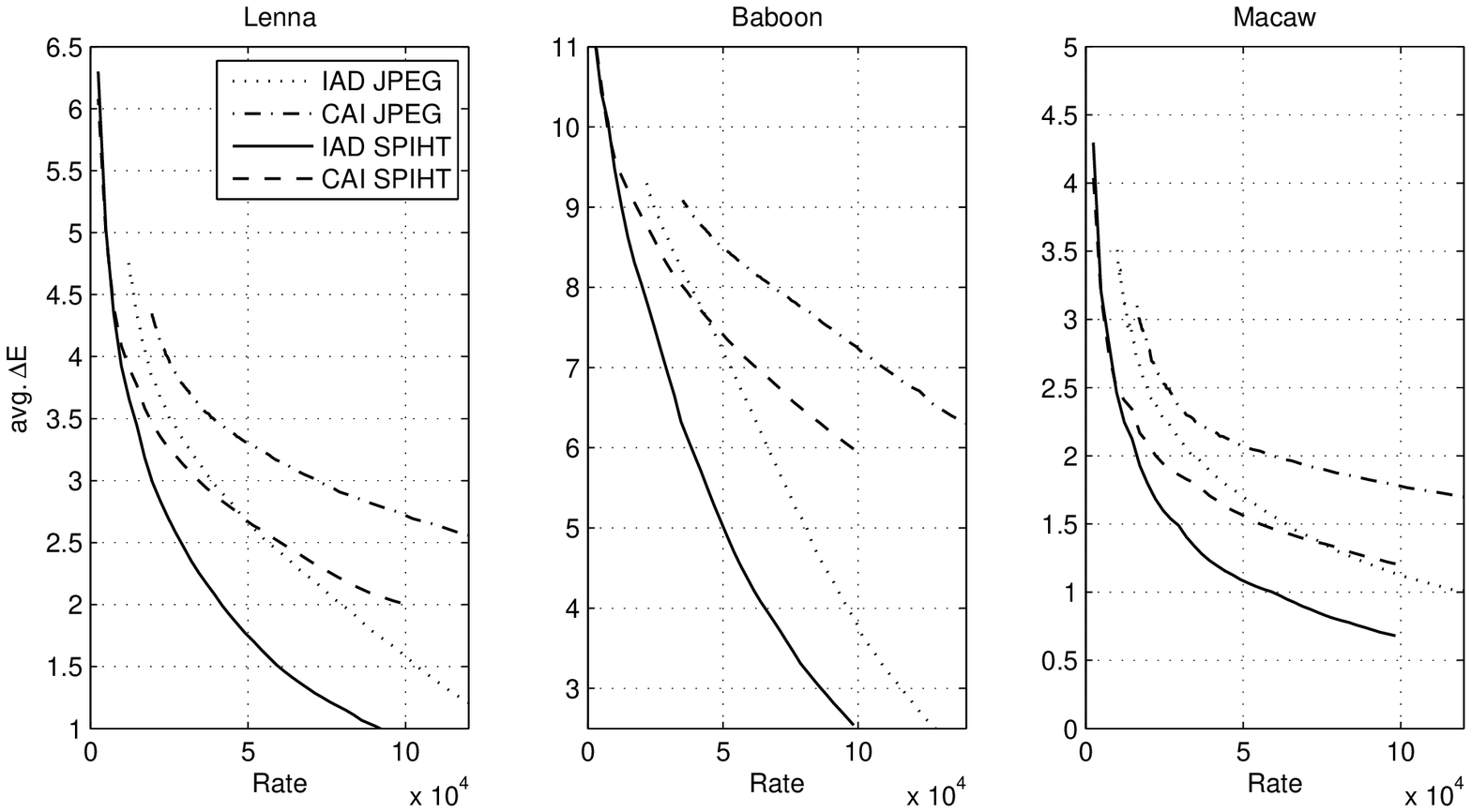}
\caption{The curves indicate the average $\Delta{E}$ of IAD and
CAI methods with bilinear interpolation.} \label{fig:error_bi}
\end{figure*}

In order to compare the error in a perceptually uniform color
space, average $\Delta{E}$ in the CIELAB space is used for a
second measure. As shown in Fig.~\ref{fig:error_bi}, IAD methods
(rotation with 64 by 64 block transform) provide smaller errors
than CAI methods with both JPEG (under all compression ratios
considered) and SPIHT (up to more than $50:1$).

In addition to the higher PSNR gain, lower average $\Delta{E}$ and
lower complexity, the IAD algorithms have other advantages such as
lower blocking artifacts after JPEG coding and fast consecutive
capturing. Reduction of blocking artifacts is achieved because a
lower compression ratio is used at a given rate. Additionally,
because interpolation is done after decompression, it can reduce
blocking artifacts similar to a de-blocking processing after JPEG
decompression. Moreover luminance data and chrominance data use
different block shapes, which may also help to reduce blocking
artifacts. Figs. \ref{fig:lenna_comp} (b) and (c) shows the result
of CAI and IAD methods, respectively. As expected, the result of
the CAI method shows more blocking artifacts. Finally, fast
consecutive capturing is possible since the compression time is
shorter because only around half of Y data has to be encoded,
while interpolation (in case of 2 by 2 format conversion) and post
processing functions are not needed during the capture process.

\section{Comparison with adaptive interpolation}
\label{sec:adaptive} In the proposed methods, interpolation is
done after decompression, so more complex interpolation can be
applied without increasing the complexity of the encoding system.
Therefore, with lower complexity than that of bilinear
interpolation, we can achieve visually better results. Since
coding errors can give negative effects to the performance of
interpolation, in this chapter, we compare the performance of
proposed methods with more complex interpolation methods.

From the compression viewpoint, bilinear interpolation is a good
method because it results in smoother (and thus easier to
compress) images. Also in the IAD algorithms, the color components
generated by using the bilinear interpolation have an error that
results from averaging the error of neighbor pixels, so that the
average distortion of interpolated color components can be lower
than that of coded color components (similar to the 1-D case shown
in (\ref{eq:MSEI})). This is also confirmed by the experimental
results shown in Fig.~\ref{fig:block} (a) and
Fig.~\ref{fig:final_result} (c) (SPIHT). Note that the two figures
have different horizontal axis and the rate used in
Fig.~\ref{fig:final_result} is 1.5 times larger than that of
Fig.~\ref{fig:block}. The result verifies that PSNR is increased
after interpolation except at high bit-rates (where round-off
error plays an important role because the coding errors are
relatively small).

Although bilinear interpolation is simple and fast, it works like
a low pass filtering and does produce smoothing of edges. To
preserve more edge information, several different adaptive
interpolation algorithms have been proposed \cite{Gunturk05SP}.
Depending on the local information, adaptive interpolation
algorithms take a different interpolation method and use the
correlation of different color components. After applying the
adaptive interpolation, the interpolated image has more edge
information (i.e., more high frequency components) and it cannot
be easily compressed. In this sense the IAD algorithms have an
advantage. Note that the IAD algorithms perform interpolation
after decoding, so the coded data is independent of interpolation
algorithms. But due to lossy compression, IAD and CAI algorithms
have different data before the interpolation. Therefore it could
happen that they have different edge information and take a
different directional interpolation method for pixels at the same
position. This results in high distortion in the generated color
components. Also, the error of one color component is involved in
the interpolation of other color components and the distortion of
generated pixels can be increased. As a result, by using the
adaptive interpolation, the IAD algorithms achieve some gains from
data smoothness before compression (especially in the case of
rotation with 64 by 64 block format conversion) but may lose in
performance from choosing different directions during
interpolation due to distorted data. Therefore when
error-sensitive interpolation methods (which are very sensitive
about the quantization noise of existing compression methods) are
required, interpolation aware compression methods (which can keep
more information used in the interpolation) are needed. But in
this paper, we mainly focus on the performance comparison between
IAD and CAI algorithms, and we use existing compression methods
with minor modification.

To verify the performance of the IAD algorithms with the adaptive
interpolation, we consider 3 different adaptive interpolation
algorithms, namely, constant hue-based, gradient based and
median-based interpolation~\cite{Ramanath02JEI}.

Constant hue-based interpolation is proposed by Cok~\cite{Cok87P}
and Kimmel~\cite{Kimmel99IP}, where hue is defined by a vector of
ratios as ($R/G,~B/G$). In this algorithm, the green color
component is used as a denominator and a small error of the green
component may induce a large error in hue, especially when green
values are small. Therefore the IAD algorithms do not provide good
performance when this interpolation is applied.

\begin{figure*}[tb]
\centering
 \begin{tabular}{cc}
   \includegraphics[width=7.8cm]{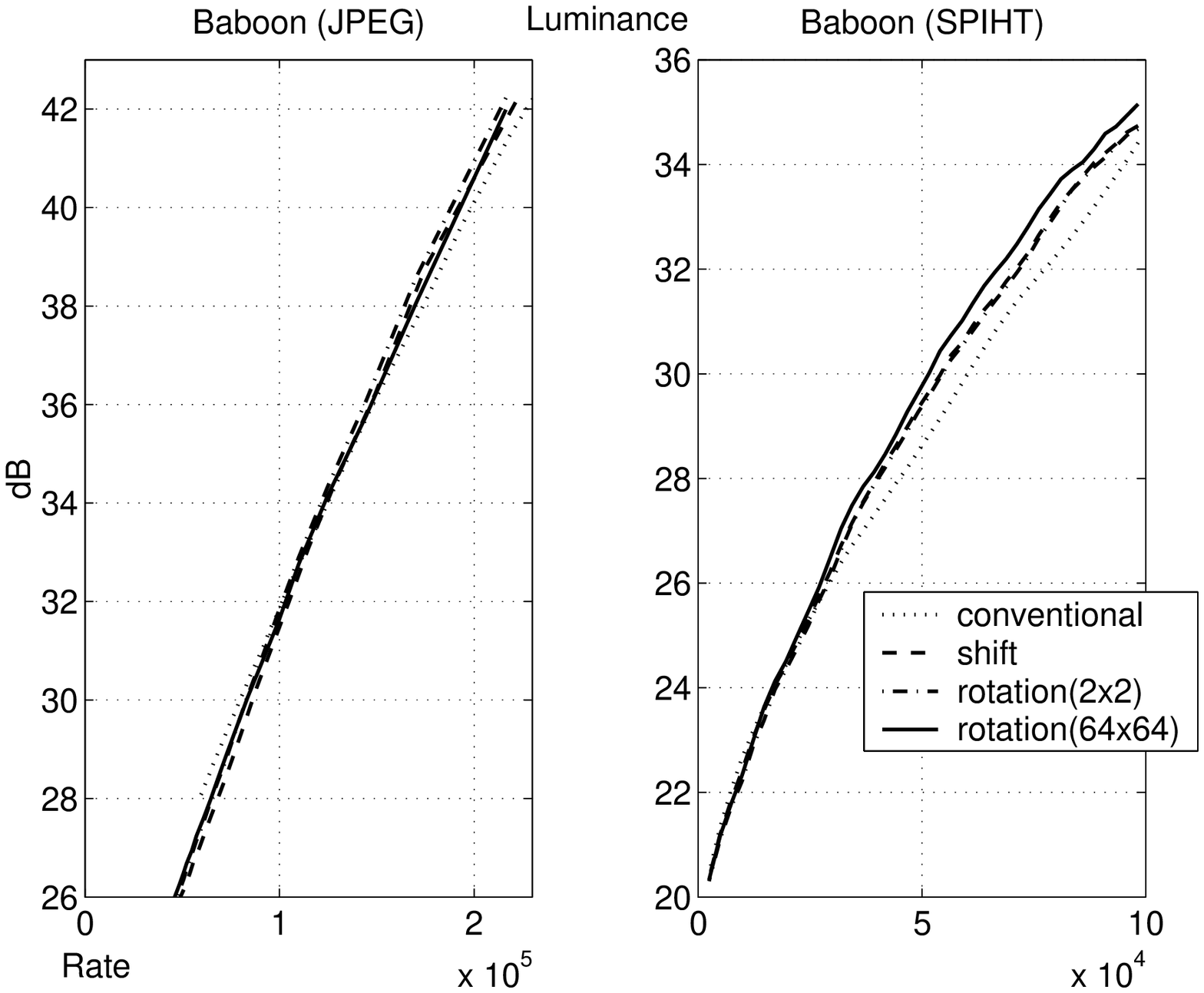} &
   \includegraphics[width=7.8cm]{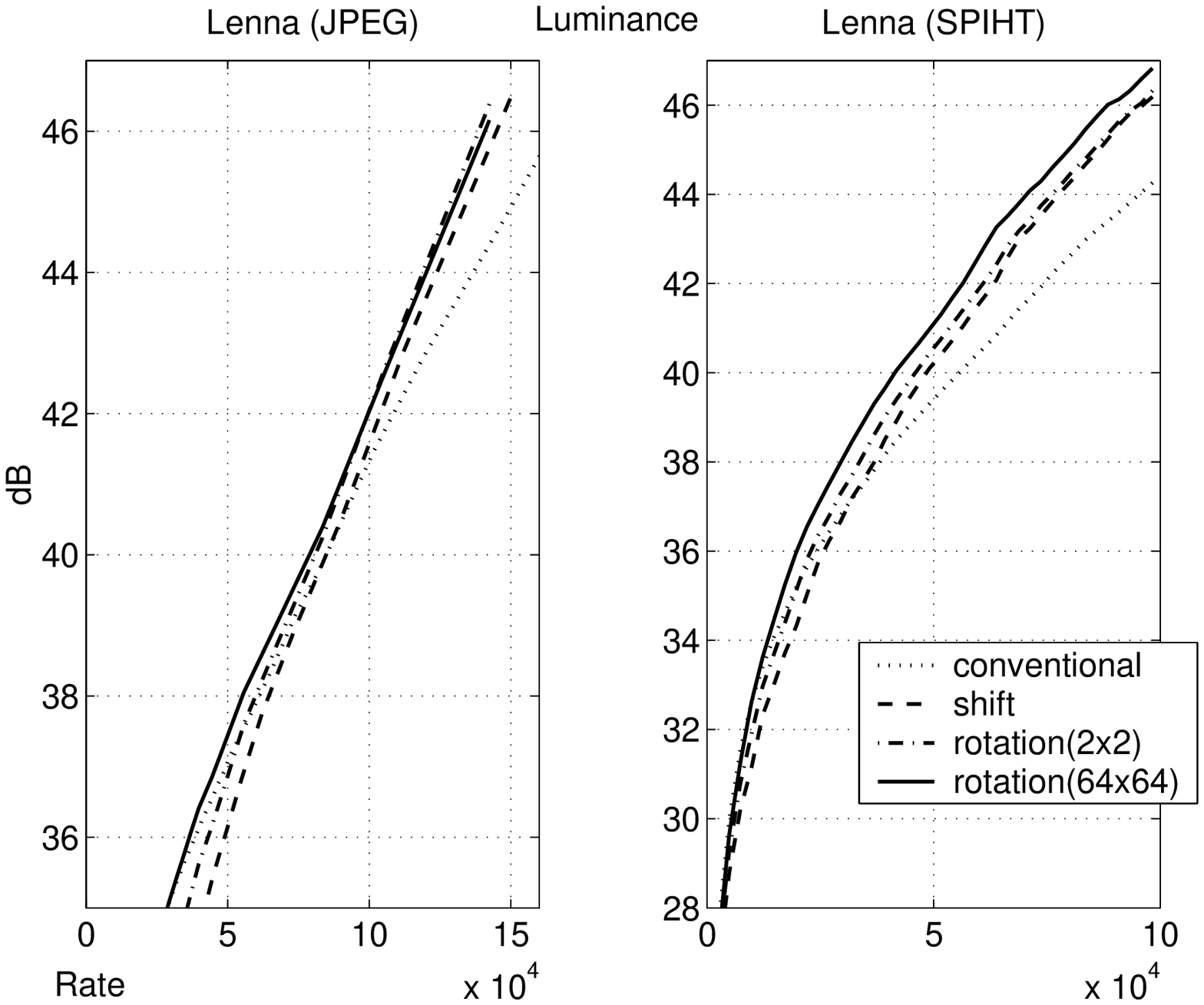} \\
   (a)&(b)\\
   \includegraphics[width=7.8cm]{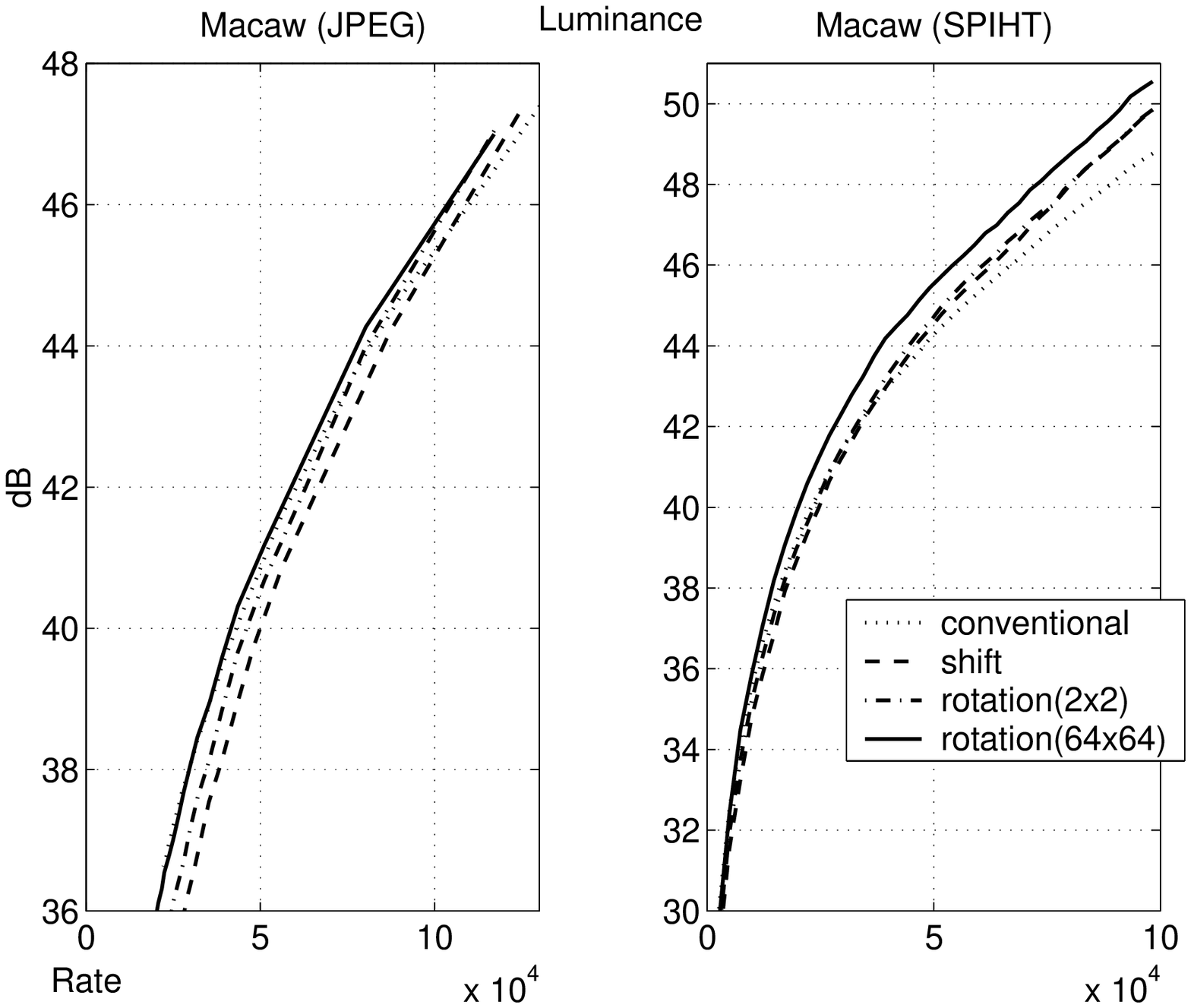} &
   \includegraphics[width=7.8cm]{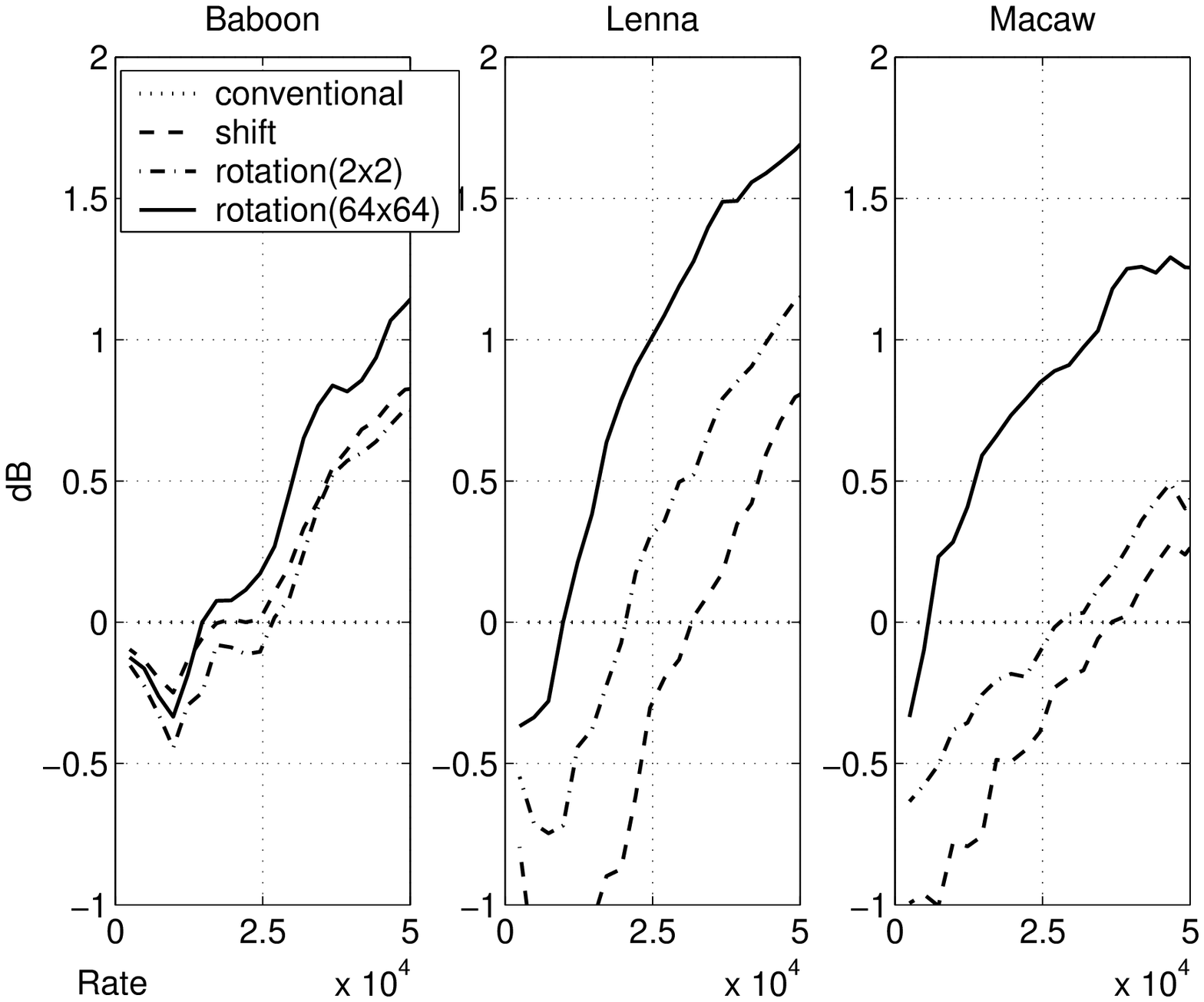} \\
   (c)&(d)\\
  \end{tabular}
\caption{The curves in (a), (b) and (c) indicate the luminance
PSNR after applying overall coding schemes with gradient based
interpolation. The curve in (d) indicates the PSNR gain of
different IAD methods against the CAI method with SPIHT. }
\label{fig:final_result_laro}
\end{figure*}

Gradient based interpolation is proposed by Laroche and
Prescott~\cite{Laroche94P}. In this algorithm, at first, green
components on blue (red) pixel positions are determined by using
directional bilinear interpolation, where the direction is
selected by the gradient of neighboring blue (red) components.
After determining green components, blue (red) components are
interpolated from the differences between blue (red) and green
components. Fig.~\ref{fig:final_result_laro} shows the coding
results of IAD algorithms. With JPEG compression, the performance
of IAD algorithms (except rotation with 64 by 64 block format
conversion) is worse than that of the CAI algorithm since
different direction is determined by large error in high frequency
components and blocking effect and the error of green components
also affects red and blue components. But with SPIHT compression,
the coding error is evenly distributed and different directional
interpolation is reduced. Therefore as shown in
Fig.~\ref{fig:final_result_laro} (d), the IAD algorithms
outperform the CAI algorithm although the gain is smaller than
when bilinear interpolation is used (shown in
Fig.~\ref{fig:diff_l}).

\begin{figure*}[tb]
\centering
 \begin{tabular}{cc}
   \includegraphics[width=7.8cm]{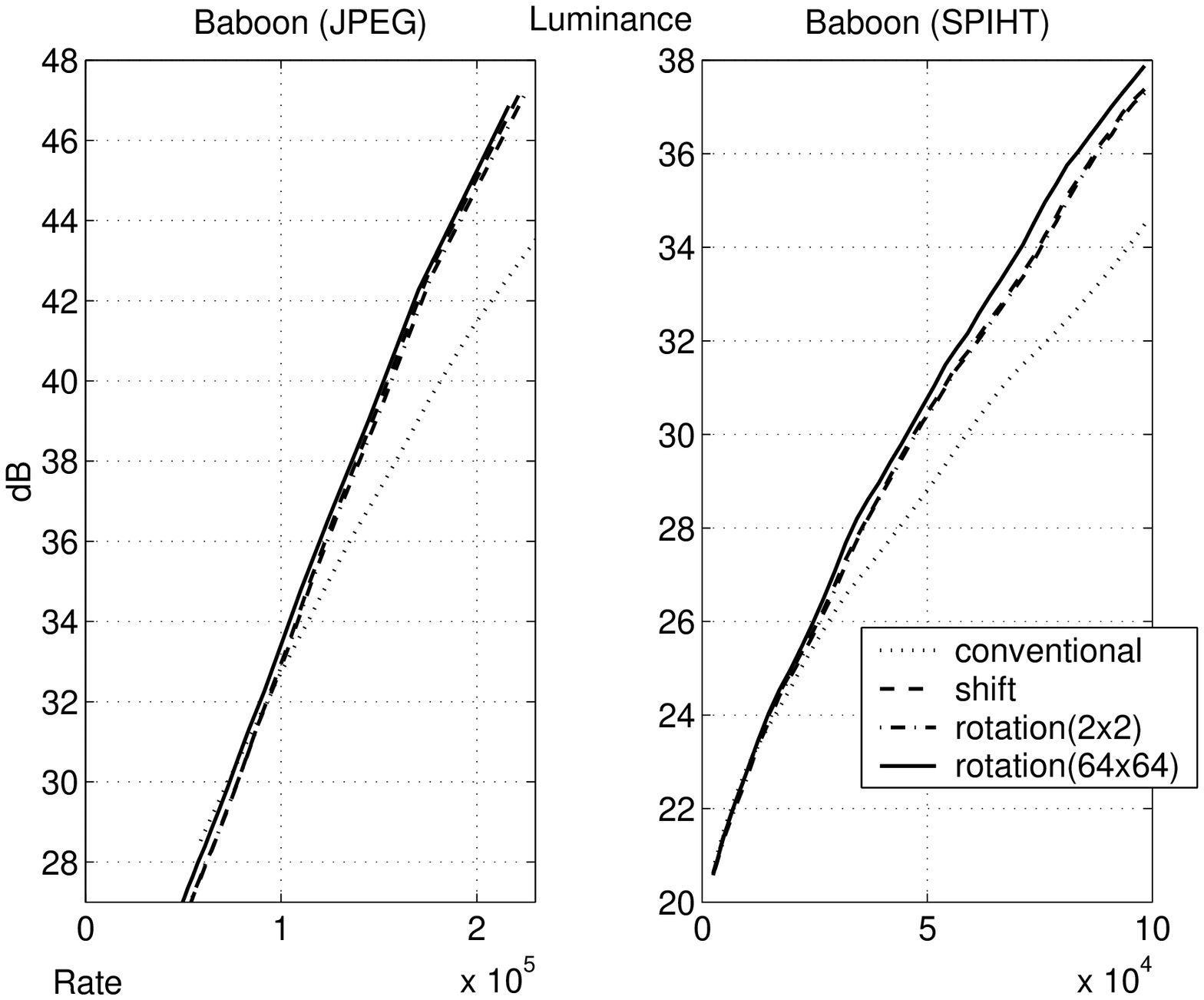} &
   \includegraphics[width=7.8cm]{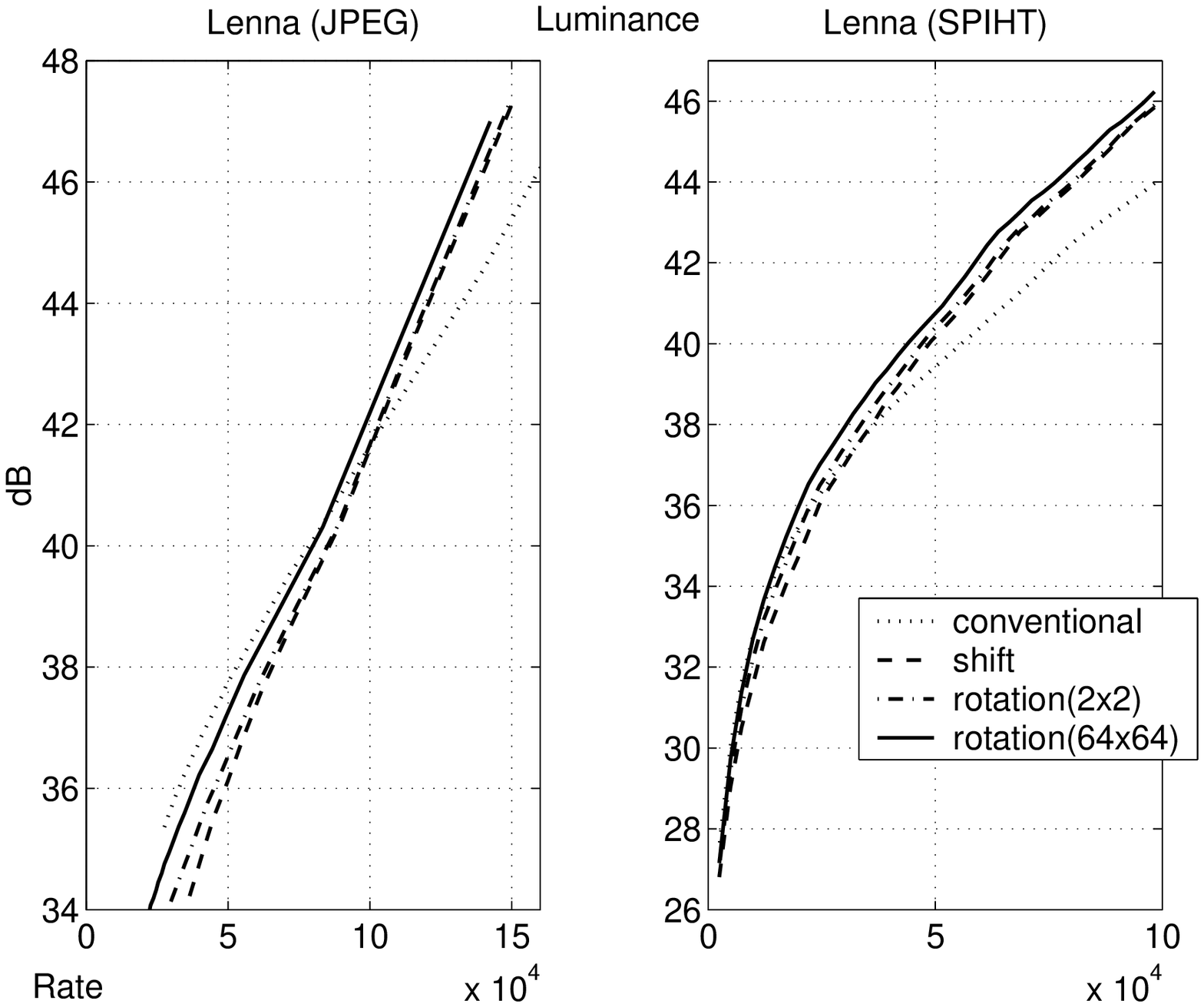} \\
   (a)&(b)\\
   \includegraphics[width=7.8cm]{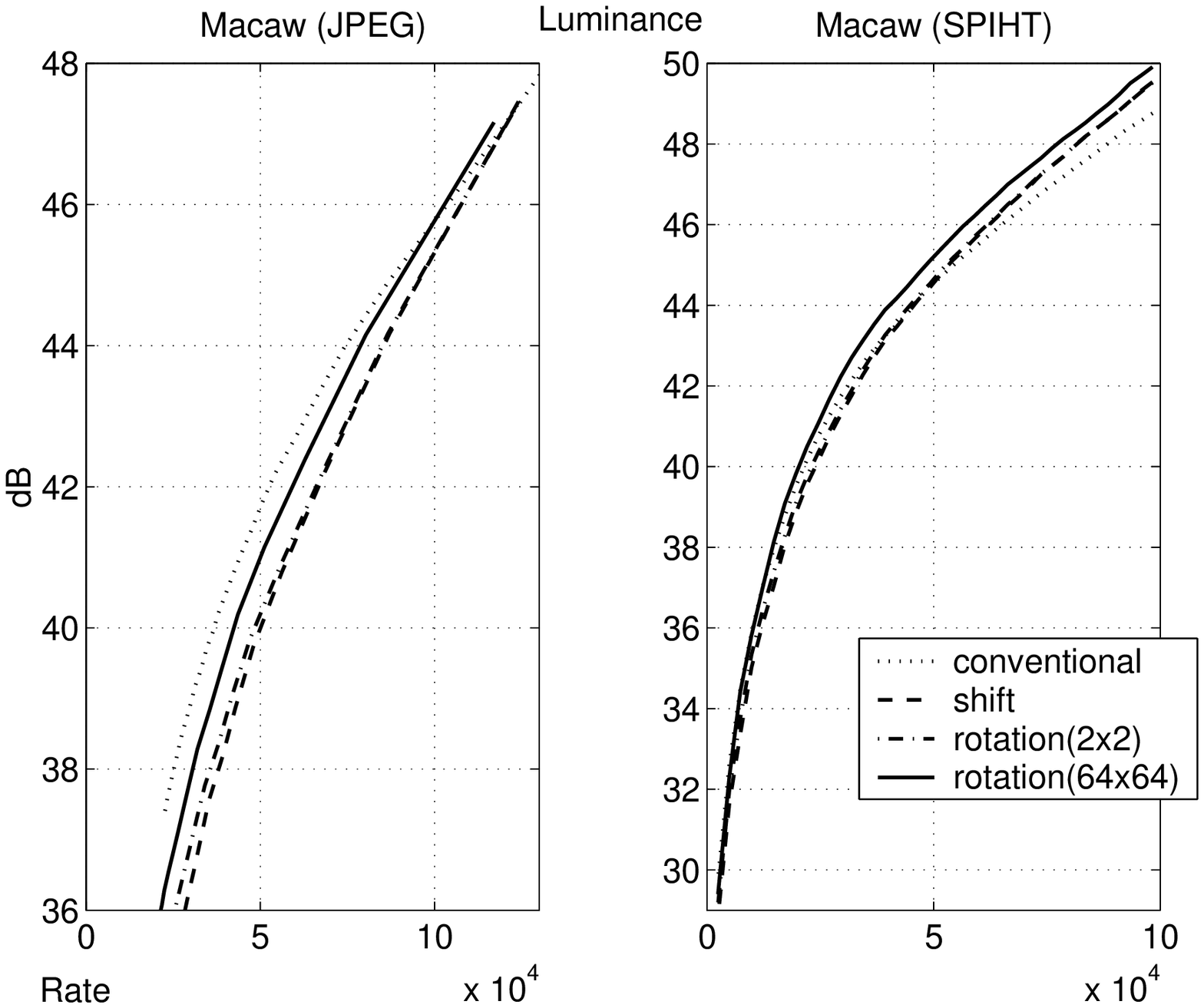} &
   \includegraphics[width=7.8cm]{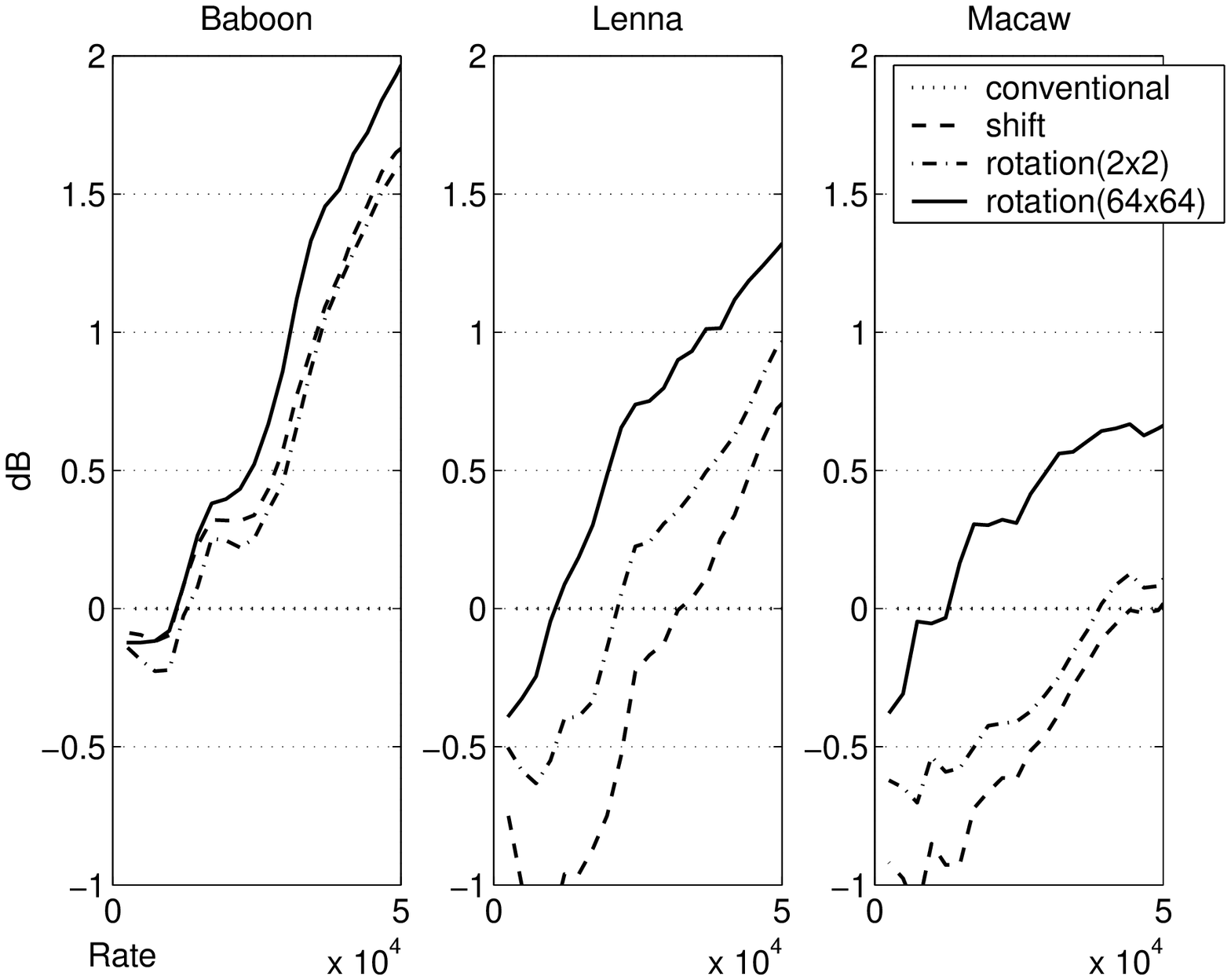} \\
   (c)&(d)\\
  \end{tabular}
\caption{The curves in (a), (b) and (c) indicate the luminance
PSNR after applying overall coding schemes with median-based
interpolation and SPIHT. The curve in (d) indicates the PSNR gain
of different IAD methods against the CAI method with SPIHT.}
\label{fig:final_result_free}
\end{figure*}

\begin{figure*}[tb]
\centering
\includegraphics[width=15cm]{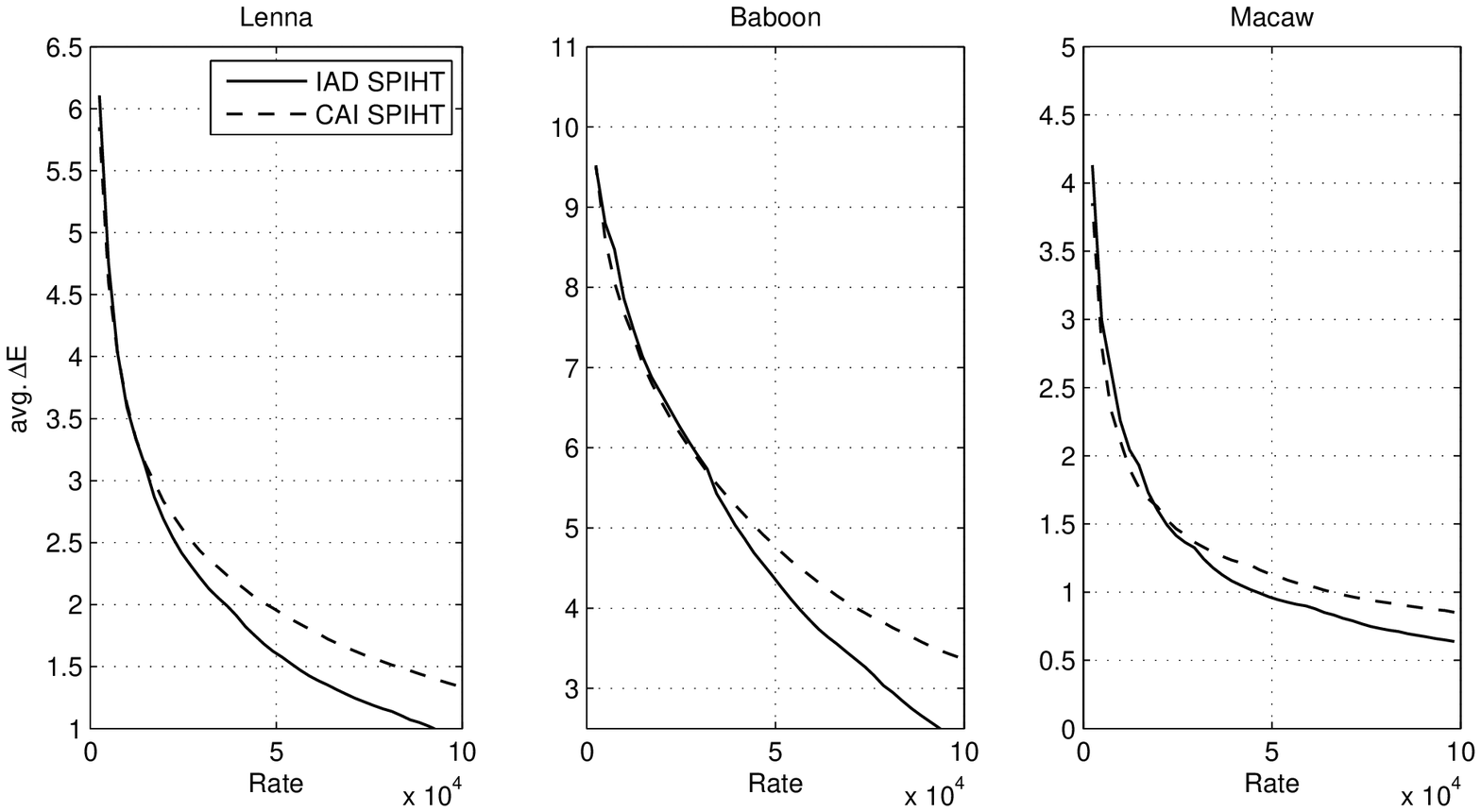}
\caption{The curves indicate the average $\Delta{E}$ of IAD and
CAI methods with median-based interpolation and SPIHT.}
\label{fig:error_free} \vspace{10cm}
\end{figure*}

\begin{figure}[tb]
\begin{minipage}[b]{1.0\linewidth}
\centering
\includegraphics[angle=-90,width=16.5cm]{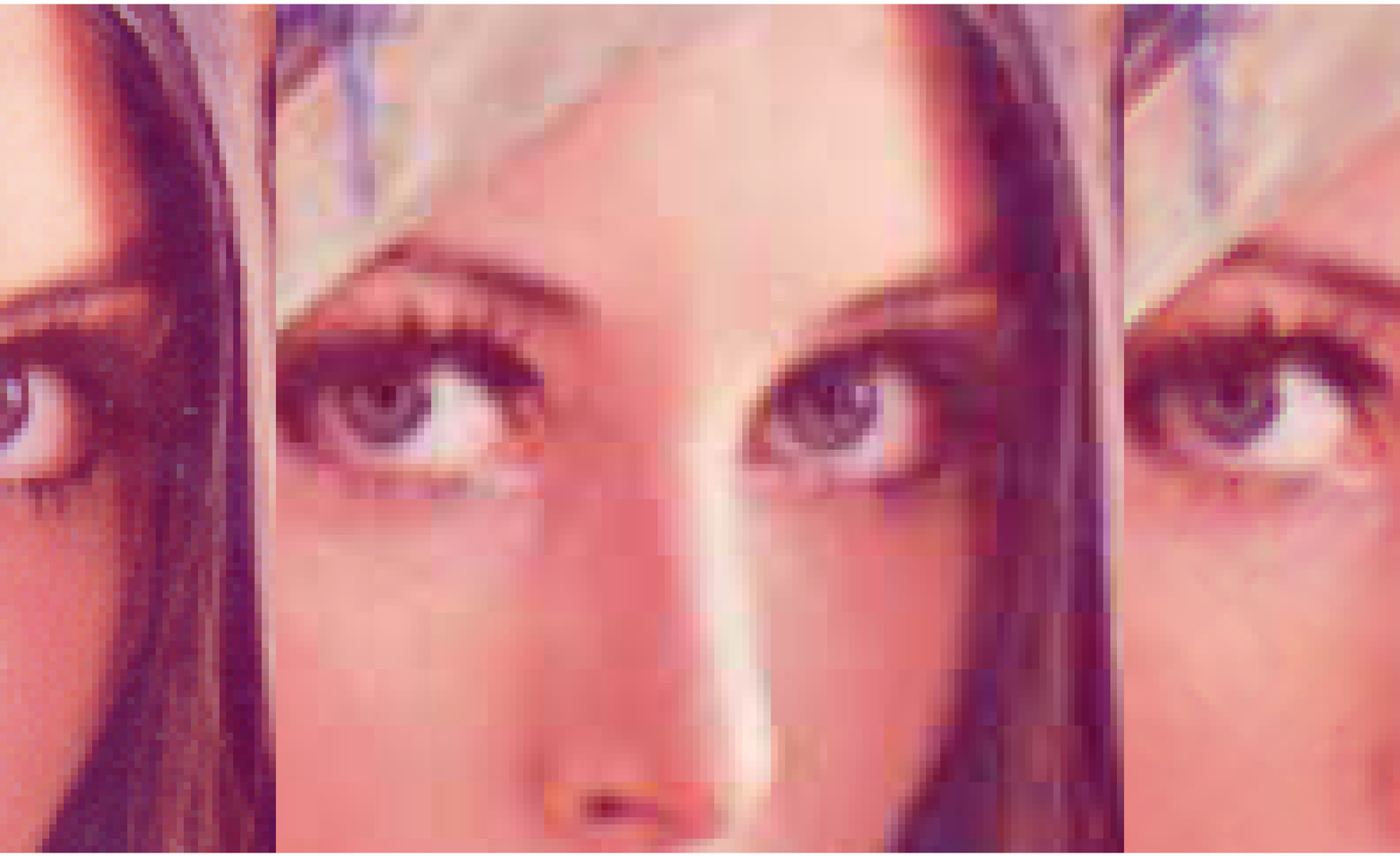}
\centerline{(a) \hspace{4.9cm} (b) \hspace{4.9cm} (c)}
\end{minipage}
\caption{Comparison of blocking artifacts. (a) is the original
image, and (b) and (c) are the images after applying CAI and IAD,
respectively. Bilinear interpolation and JPEG compression are
used, where the compression ratio is $35.2:1$.}
\label{fig:lenna_comp}
\end{figure}

\begin{figure}[tb]
\begin{minipage}[b]{1.0\linewidth}
\centering
\includegraphics[angle=-90,width=16.5cm]{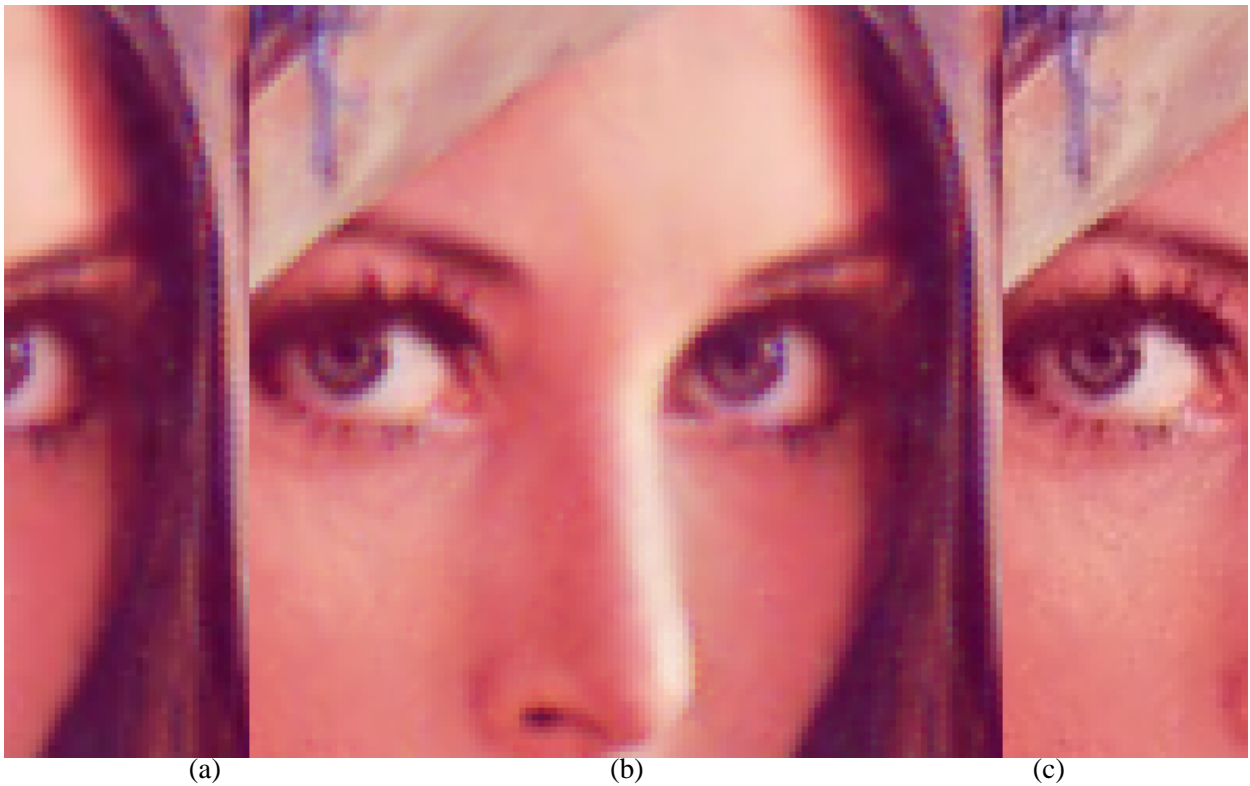}
\centerline{(a) \hspace{4.9cm} (b) \hspace{4.9cm} (c)}
\end{minipage}
\caption{Comparison of different interpolation. (a) and (b) are
the images after applying CAI and IAD, respectively. Bilinear
interpolation and SPIHT compression are used. (c) is the images
after applying CAI with Freeman interpolation. Compression ratio
used is $16:1$. Note that the encoding of (b) and (c) are
identical.} \label{fig:lenna_comp2}
\end{figure}

The performance is also tested with Median-based interpolation
(proposed by Freeman~\cite{Freeman88P}) which employs two step
processes. The first pass is the bilinear interpolation and the
second pass is selecting the median of color differences of
neighboring pixels. Fig.~\ref{fig:final_result_free} shows the
coding results of IAD algorithms with a 3 by 3 median filter.
Similar to the gradient-based interpolation, the IAD algorithms
provide worse results when JPEG is applied. But with SPIHT, IAD
algorithms still provide better results up to $20:1$ or $40:1$
compression ratio depending on the format conversion methods.
Fig.~\ref{fig:error_free} shows average $\Delta{E}$ in the CIELAB
space. IAD algorithms provide better results up to more than
$20:1$ compression ratio and then the average $\Delta{E}$ of both
algorithm goes similar. Fig.~\ref{fig:lenna_comp2} shows the
results after applying bilinear and Freeman interpolation. As
expected, The image with Freeman interpolation
(Fig.~\ref{fig:lenna_comp2} (c)) is sharper and close to the
original image (Fig.~\ref{fig:lenna_comp} (a)).

As a result, the IAD algorithms with SPIHT provide better results
with the gradient based and median-based interpolation. But due to
coding inefficiency, irregular coding error and blocking effect,
adaptive interpolation for a given pixel may be different
depending on using quantized or unquantized data, resulting in
potential degradation after interpolation when quantized data is
used interpolation for each pixel. Therefore the performance of
the IAD algorithms with JPEG is worse.

\section{CONCLUSION}
\label{sec:capt_conclusion} \vspace{0.0cm} In this paper, we
investigated the redundancy decreasing method that merges an image
processing stage and an image compression stage. Several color
format conversion algorithms and shift and rotation transforms are
introduced to compress CFA images before making full color images
by interpolation. We showed that the proposed algorithms
outperform the conventional method in the full range of
compression ratios for JPEG coding with bilinear interpolation and
up to $20:1$ or $40:1$ compression ratio (depending on the color
format conversion and interpolation methods) for SPIHT coding when
the bilinear, gradient based and median-based interpolation are
applied. Also we analyzed the PSNR gain and explained why it
becomes higher as the compression ratio becomes lower, showing
also a 1D DPCM sequence example to provide some intuition. Because
the proposed algorithms use only around half the amount of Y data
and only need an additional simple transform, the computational
complexity can be decreased. Also adaptive interpolation methods
can be applied without increasing the encoder complexity,
 so fast consecutive capturing can be achieved with visually better results.

In this paper, we tried to minimize the changes to existing
compression methods in order to focus on the performance of
changing the encoding order (i.e., the order of interpolation and
compression). The performance of the IAD method can be improved
with different entropy coding (in SPIHT). Also, in the rotation
method, a new quantization table may be useful for a JPEG codec,
due to the directional difference of human visual sensitivity.

\bibliographystyle{IEEEtran}

\end{document}